\documentclass{elsart}
\usepackage{graphicx}
\usepackage{amssymb}

\newcommand{\be}{\begin{equation}}
\newcommand{\ee}{\end{equation}}
\newcommand{\bn}{$ \begin{array}{l} \vspace{-0.2cm}}
\newcommand{\en}{\end{array} $}

\newcommand{\Msun}{\mbox{$M_{\odot}\;$}}
\newcommand{\Msunend}{\mbox{$M_{\odot}$}}
\newcommand\appropto{\buildrel \sim \over \propto}

\newcommand{\egth}{{\rm G}_{300}}
\newcommand{\egthpi}{{\rm G}^\pi_{300}}

\newcommand{\kFn}{{k_{F_n}}}

\newcommand{\mevt}{{\rm MeV~fm}^{-3}}
\newcommand{\gcmt}{{\rm g~cm}^{-3}}
\newcommand{\kFp}{k_{F_p}}
\newcommand{\kFe}{k_{F_e}}
\newcommand{\msun}{{M_{\odot}}}
\newcommand{\ecrusti}{\epsilon_{\rm crust}}

\def\ls{\lower0.5ex\hbox{$\; \buildrel < \over \sim \;$}}

\newcommand{\gsim}{\stackrel{\textstyle >}{_\sim}}

\begin{document}

\begin{frontmatter}

\title{The Cooling of Compact Stars}
\protect\vspace{-1cm}
\author[Mexico]{Dany Page$^1$}\ead{page@astroscu.unam.mx},
\author[Garching]{Ulrich Geppert}\ead{urme@xray.mpe.mpg.de},
\author[SanDiego]{Fridolin Weber$^2$}\ead{fweber@sciences.sdsu.edu}

\address[Mexico]{Departamento de Astrof\'{\i}sica Te\'orica, Instituto
                 de Astronom\'{\i}a, \\ Universidad Nacional
                 Aut\'onoma de M\'exico, 04510 Mexico D.F.}

\address[Garching]{Max-Planck-Institut f\"ur extraterrestrische Physik, \\
                   Giessenbachstrasse, PF 1312, 85741 Garching, Germany}

\address[SanDiego]{Department of Physics, San Diego State University, \\
                  5500 Campanile Drive, San Diego, California 92182, USA}

\thanks{D. P. work is partially supported by a grant from UNAM-DGAPA,
        \#IN112502}

\thanks{F. W. research was supported by an award from Research
        Corporation.}

\protect\vspace{-0.5cm}
\begin{abstract}
The cooling of a compact star depends very sensitively on the state of
dense matter at supranuclear densities, which essentially controls the
neutrino emission, as well as on the structure of the stellar outer
layers which control the photon emission.  Open issues concern the
hyperon population, the presence of meson condensates, superfluidity
and superconductivity, and the transition of confined hadronic matter
to quark matter.  This paper describes these issues and presents
cooling calculations based on a broad collection of
equations of state for neutron star matter and strange matter.  These
results are tested against the body of observed cooling data.
\end{abstract}

\end{frontmatter}

\protect\vspace{-1.0cm}

\section{Introduction}
\label{intro}

Cooling simulations, confronted with soft X-ray, extreme
UV, UV and optical observations of the thermal photon flux emitted
from the surfaces of neutron stars, provide most valuable information
about the physical processes operating in the interior of these objects.
The predominant cooling mechanism of hot (temperatures $T\gsim
10^{10}$~K) newly formed neutron stars is
neutrino emission, with an initial cooling time scale of seconds.
Neutrino cooling still dominates for at least the first thousand years, 
and typically for much longer in slow (standard) cooling scenarios. 
Photon emission eventually overtakes neutrinos when the internal 
temperature has sufficiently dropped.
Being sensitive to the adopted
nuclear equation of state (EOS), the stellar mass, the assumed
magnetic field strength, superfluidity, meson condensates, and the
possible presence of color-superconducting quark matter, theoretical
cooling calculations serve as a principal window on the properties of
super-dense hadronic matter and neutron star structure.  The thermal
evolution of neutron stars also yields information about such
temperature-sensitive properties as transport coefficients, crust
solidification, and internal pulsar heating mechanisms.

We will present here an overview of the current status of neutron star
cooling calculations tested against the steadily growing body of
observed cooling data on neutron stars.  The reader can also find a
complementary approach in Ref.~\cite{YP04}.  Space limitation forbid
us to dwell into the discussion of observational data and we refer the
reader to Ref.~\cite{PLPS04} for a recent compilation
 and to two recent observational reviews \cite{PZ03,H04}
to get a flavor of the difficulties involved in the
analysis and interpretation of the data.

The paper is organized as follows.  In \S~\ref{sec:basic_eqs} we
introduce the basic equations and the physics input that governs
neutron star cooling. Some simple analytical solutions to the cooling
equations are presented in \S~\ref{sec:analytical}. The minimal
cooling paradigm, which assumes that no enhanced neutrino emission is
allowed in neutron stars, is presented in \S~\ref{sec:minimal}. In
\S~\ref{sec:enhanced} we discuss enhanced neutron star cooling via the
direct Urca process, meson condensates, and quarks. The cooling
behavior of compact stars made of absolutely stable strange quark
matter is explored in \S~\ref{sec:ss}. The impact of magnetic fields
on cooling and heating mechanisms are discussed in
\S~\ref{sec:magcrust} and \S~\ref{sec:heating}, respectively while
\S~\ref{sec:SXRT} considers cooling neutron stars in transiently
accreting binary systems.  Conclusions are offered in
\S~\ref{sec:epilogue}.

\section{Basic Equations and Physics Input}
\label{sec:basic_eqs}


The basic features of the cooling of a neutron star are easily grasped
by simply considering the energy conservation equation for the star.
In its Newtonian formulation this equation reads
\be
\frac{dE_\mathrm{th}}{dt} = C_\mathrm{v} \frac{dT}{dt}
                   = -L_\nu - L_\gamma + H \, ,
\label{equ:energy-conservation}
\ee
where $E_\mathrm{th}$ is the thermal energy content of the star, $T$ its
internal temperature, and $C_\mathrm{v}$ its total specific heat. The energy
sinks are the total neutrino luminosity $L_\nu$, described in
\S~\ref{sec:neutrinos}, and the surface photon luminosity $L_\gamma$,
discussed in \S~\ref{sec:envelope}.  The source term $H$ includes all
possible ``heating mechanisms'' which, for instance, convert magnetic
or rotational energy into heat as summarized in \S~\ref{sec:heating}.
Some simple analytical solutions to
Eq.~(\ref{equ:energy-conservation}) will be presented in
\S~\ref{sec:analytical}.

The dominant contributions to $C_\mathrm{v}$ come from the core, constituting 
more than 90\% of the total volume, whose constituents are quantum liquids of leptons,
baryons, mesons, and possibly deconfined quarks at the highest densities.
When baryons, and quarks, become paired, as briefly described in
\S~\ref{sec:pairing}, their contribution to $C_\mathrm{v}$ is strongly
suppressed at temperatures $T \ll T_\mathrm{c}$ ($T_\mathrm{c}$ being the corresponding
critical temperature).  The crustal contribution is in principle
dominated by the free neutrons of the inner crust but, since these are
certainly extensively paired, practically only the nuclear lattice and
electrons contribute.  Extensive baryon, and quark, pairing can thus
significantly reduce $C_\mathrm{v}$, but by at most a factor of order ten since
the leptons do not pair.

All results presented in this chapter were obtained with numerical
codes which exactly solve both the energy conservation and the heat
transport equations in their General Relativistic forms (see, e.g.,
\cite{PLPS04,SWWG96}).  Since the density and chemical composition of
the star, after the proto-neutron star phase, do not change with time, 
the Tolman-Oppenheimer-Volkoff
equation of hydrostatic equilibrium is solved initially, for a given
EOS, and only the thermal equations are evolved with time.  The outermost,
low density, layers of the star do see their structure evolve during
the cooling and they are hence treated separately as an envelope which
is the outer boundary condition and is described in \S~\ref{sec:envelope}.

\subsection{The Equation of State (EOS)} 
           \label{sec:EOS}

The cross section of a neutron star can be split roughly into three, possibly
four, distinct regimes. The first regime is the star's outer crust, which
consists of a lattice of atomic nuclei and a Fermi liquid of
relativistic, degenerate electrons. The second regime, known as the
inner crust and where free neutrons appear, extends from neutron drip density, 
$\sim 4-8 \times 10^{11}~\gcmt$, to a transition density of about 
$2\times 10^{14}~\gcmt$.  Beyond that density one enters the star's third
regime, its core matter where all atomic nuclei have dissolved into
their constituents, protons and neutrons. Furthermore, because of the
high Fermi pressure, the inner core may be expected to contain baryon resonances, 
boson condensates, hyperons, or/and a mixture of deconfined up, down and strange
quarks \cite{glen97:book,weber99:book}. 
The EOS of the outer and inner crust is rather well known. 
This is very different for the EOS of the star's core which is only very poorly 
understood. Models derived for it fall into non-relativistic EOS and relativistic, 
field theoretical ones.

\begin{figure}[tb] 
\begin{center}
\includegraphics[scale=0.4,angle=-90]{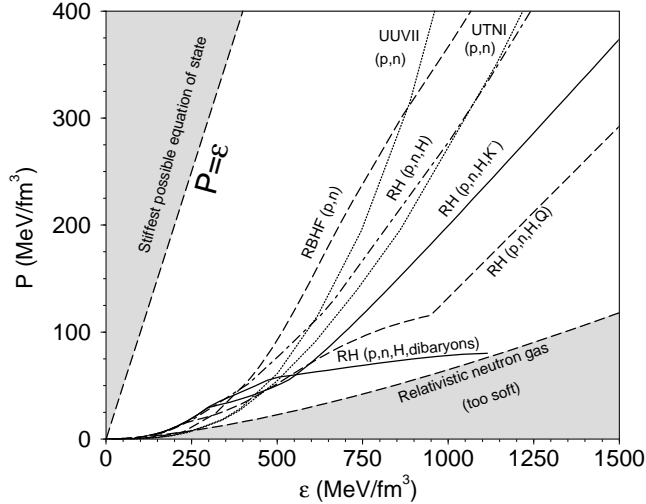}
\caption{Models for the EOS of high-density neutron star matter computed for
different compositions and many-body techniques. (p,n denote protons
and neutrons; H, K$^-$, Q stand for hyperons, K$^-$ condensate, and
quarks, respectively.)}
\label{fig:eos}
\end{center}
\end{figure}

The most frequently used non-relativistic models for the EOS are based
on the hole-line expansion (Brueckner theory), the coupled cluster
method, self-consistent Green functions, the variational approach
(Monte Carlo techniques), the semi-classical Thomas-Fermi method, and
the density functional approach based on Skyrme effective interactions
(for an overview of these methods and additional references, see for
instance Refs.\
\cite{heiselberg00:a,muether00:a,strobel97:a}). 
The forces between hadrons are described in terms of phenomenological
nucleon-nucleon interactions, possibly supplemented with three-nucleon
interactions to achieve a better agreement with the binding energy of
nuclear matter at the empirical saturation density.  Figure~\ref{fig:eos} 
shows two  sample models for the EOS of
neutron star matter (protons and neutrons only) which are obtained from
variational calculations based on the Urbana $V_{14}$ two-nucleon
interaction supplemented with the three-body interactions UVII (left
curve) and TNI (right curve) \cite{wiringa88:a}.
 
Relativistic, field-theoretical approaches to the EOS are based on
model Lagrangians which generally describe baryons as Dirac particles
interacting via the exchange of mesons.  The most important mesons are
the $\sigma$ and the $\omega$, which are responsible for nuclear
binding, while the $\rho$ meson is required to obtain the correct value for
the empirical symmetry energy. Nonlinear $\sigma$ terms need to be
included at the mean-field (relativistic Hartree, RH) level in order
to obtain the empirical incompressibility of nuclear matter
\cite{glen97:book}. Such terms are not necessary if the field
equations are solved for the relativistic Brueckner-Hartree-Fock
(RBHF) approximation
\cite{weber99:book}. In contrast to RH, where the parameters of the
theory are adjusted to the properties of infinite nuclear
matter, the RBHF method makes use of one-boson-exchange potentials
whose parameters are adjusted to the properties of the deuteron and to
the nucleon-nucleon scattering data.  Figure~\ref{fig:eos} shows
several sample EOSs computed for RH and RBHF, assuming different
particle compositions of neutron star matter
\cite{weber99:book}. Finally we mention that in
 recent years a new class of effective field theories was developed
which treat the meson-nucleon couplings as density dependent. These field
theories provide a very good description of the properties of nuclear
matter, atomic nuclei as well as neutron stars
\cite{lenske95:a,fuchs95:a,ban04:a}.

\begin{figure}[tb]
\begin{center}
\includegraphics[scale=0.4,angle=-90]{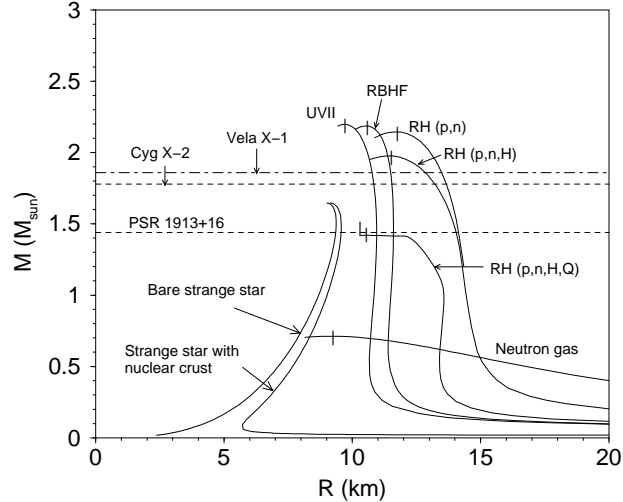}
\caption{Neutron star mass versus radius for different EOSs. The horizontal
  lines refer to the masses of Vela X-1 ($1.88\pm 0.13\, \msun$)
  \cite{quaintrell03:a}, Cyg X-2 ($1.78\pm 0.23 \, \msun$)
  \cite{orosz99:a}, and PSR 1913+16 ($1.4408\pm 0.0003\, \msun$)
  \cite{weisberg04:a} (in all three cases quoted mass uncertainties are $1\sigma$).}
\label{fig:mrad2}
\end{center}
\end{figure}

In Figure~\ref{fig:mrad2} we show the mass-radius relationships of
neutron stars for different EOSs:
it illustrates the well-known fact that this relationship is
very sensitive to the underlying model for the EOS.

\begin{figure}[tb]
\begin{center}
\includegraphics[scale=0.4]{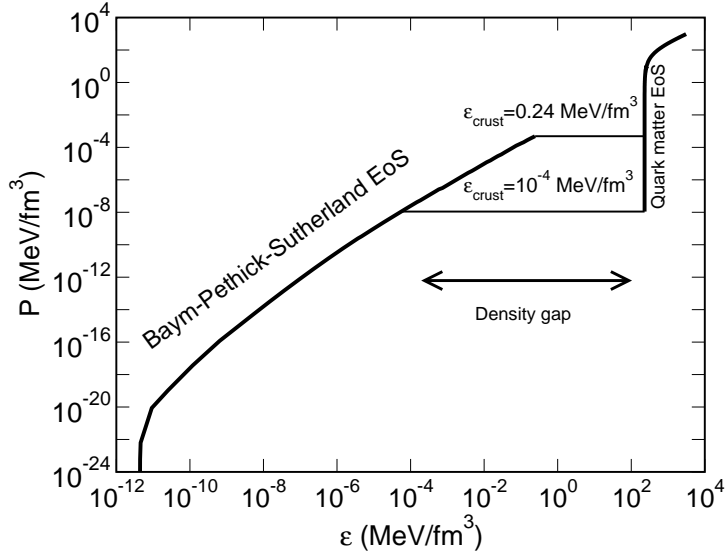}
\caption{EOS of strange quark matter surrounded by nuclear matter. The
  maximal possible nuclear matter density is determined by neutron
  drip which occurs at $\ecrusti = 0.24~\mevt$ ($4.3\times
  10^{11}~\gcmt$). Any nuclear density that is smaller than that is
  possible. As an example, we show the EOS for a sample density of
  $\ecrusti = 10^{-4}~\mevt$ ($10^8~\gcmt$).}
\label{fig:eos.ss}
\end{center}
\end{figure}

Phase transitions to boson condensates, hyperonic matter, or quark matter,
soften the EOS, hence reducing both the stars gravitational mass and radius
and resulting in a smaller maximum mass sustainable by the EOS.
A generically
different mass-radius relationship is obtained for stars made of
absolutely stable strange quark matter (strange stars). For them $M
\propto R^3$, which is only significantly modified if $M$ is close to
the mass peak.  As pointed out in
\cite{alcock86:a}, strange stars can carry a solid nuclear crust whose
density at its base is strictly limited by neutron drip.  This is made
possible by the displacement of electrons at the surface of strange
matter, which leads to a strong electric dipole layer there. The
associated electric field is so strong that it holds open a gap
between the nuclear crust and quark matter, preventing the conversion
of the crust into the hypothetical lower-lying ground state of strange
matter.  Obviously, free neutrons, being electrically charge neutral,
cannot exist in the crust because they do not feel the Coulomb barrier
and thus would gravitate toward the strange quark matter core where
they are converted, by hypothesis, into strange matter.  Consequently,
the density at the base of the crust (inner crust density) will always
be smaller than neutron drip. The situation is graphically illustrated
in Figure~\ref{fig:eos.ss} which shows the EOS of a strange star
\cite{weber99:book,glen92:crust} carrying a nuclear crust. 
Since the crust is bound to the quark matter core by gravity rather
than confinement, the mass-radius relationship of a strange star with
a nuclear crust is qualitatively similar to the one of purely
gravitationally bound neutron stars, as shown in Figure~\ref{fig:mrad2}.

\subsection{Neutrino Processes} 
           \label{sec:neutrinos}

As already mentioned in the introduction, neutron stars are born with
temperatures in excess of $10^{10}$~K.  The dominating cooling
mechanism of such objects for the first several thousand years after
birth is neutrino emission from the interior.  After that, cooling via
photon emission from the star's surface takes over.
Tables~\ref{tab:emis.core} and \ref{tab:emis.quark} summarize the
dominant neutrino emitting processes together with their efficiency
for neutron star cooling and we now briefly describe the most
important ones.  The reader is referred to
Refs.~\cite{weber99:book,P92,Yetal01,V01} for more details.  We will
nevertheless comment in \S~\ref{sec:minor-nu} on minor neutrino
emission processes which may become important in cases where all the
dominant ones, as discussed in \S~\ref{sec:pairing}, are suppressed by
pairing.

\begin{table}[t]
\begin{center}
\caption{Dominant neutrino emitting processes in neutron star cores, in absence of hyperons$^a$ and quarks$^b$.}
\label{tab:emis.core}
\begin{tabular}{llcc} 
\hline 
     Name           &               Process$^c$              &       Emissivity$^d$       &  \\ 
                    &                                        &   (erg cm$^{-3}$ s$^{-1}$) &               \\
\hline 
\parbox[c]{3.4cm}{Modified Urca cycle\\ (neutron branch)} &
\rule[-0.4cm]{0.02cm}{0.85cm}
\bn n+n \rightarrow n+p+e^-+\bar\nu_e \\ n+p+e^- \rightarrow n+n+\nu_e \en  & 
$\sim 2\!\!\times\!\! 10^{21} \: R \: T_9^8$ & Slow \\
\parbox[c]{3.4cm}{Modified Urca cycle\\ (proton branch)}  &
\rule[-0.4cm]{0.02cm}{0.85cm}
\bn p+n \rightarrow p+p+e^-+\bar\nu_e \\ p+p+e^- \rightarrow p+n+\nu_e \en  & 
$\sim 10^{21} \: R \: T_9^8$ & Slow \\
Bremsstrahlung          &
\bn n+n \rightarrow n+n+\nu+\bar\nu \\ n+p \rightarrow n+p+\nu+\bar\nu \vspace{-0.2cm} \\
    p+p \rightarrow p+p+\nu+\bar\nu \en                                     & 
$\sim 10^{19} \: R \: T_9^8$  & Slow \\
\parbox[c]{3.5cm}{Cooper pair \\ formations}          &
\bn    n+n \rightarrow [nn] +\nu+\bar\nu \\ p+p \rightarrow [pp] +\nu+\bar\nu \en  & 
\bn \sim 5\!\!\times\!\! 10^{21} \: R \: T_9^7 \\ \sim 5\!\!\times\!\! 10^{19} \: R \: T_9^7 \en & Slow \\
Direct Urca cycle        & 
\rule[-0.4cm]{0.02cm}{0.85cm}
\bn n \rightarrow p+e^-+\bar\nu_e \\ p+e^- \rightarrow n+\nu_e \en              & 
$\sim 10^{27} \: R \: T_9^6$ & Fast \\
$\pi^-$ condensate &$n+<\pi^-> \rightarrow n+e^-+\bar\nu_e$  &
$\sim 10^{26} \: R \: T_9^6$  & Fast \\
$K^-$ condensate   &$n+<K^-> \rightarrow n+e^-+\bar\nu_e$  &
$\sim 10^{25} \: R \: T_9^6$  & Fast \\
\hline 
\end{tabular}
\end{center}
\vspace*{.4cm}
\noindent
$^a$ In the presence of hyperons, most processes listed here have
     corresponding ones with hyperons replacing nucleons. \\
$^b$ See Table~\ref{tab:emis.quark} for quark processes. \\
$^c$ Where $\mu^-$ are present, all processes involving $e^-$ have a
     corresponding one with $\mu^-$, $\nu_{\mu}$, and $\bar\nu_{\mu}$
     replacing $e^-$, $\nu_e$, and $\bar\nu_e$, respectively. \\
$^d$ Quoted emissivities are only indicative: each process has its
     specific dependences on medium and particle densities, effective
     masses, plus medium effect corrections, for which we refer the
     reader to Refs.~\cite{weber99:book,P92,Yetal01,V01}; the many $R$
     factors are the respective temperature dependent control
     functions which take into account the effects of pairing as
     discussed in \S~\ref{sec:pairing} and detailed in
     Ref.~\cite{Yetal01}.
\end{table}

\noindent
{\bf Direct Urca Processes.}
The beta decay and electron capture processes among nucleons, 
$n \rightarrow p + e^- + \bar{\nu}_e$ and $p + e^- \rightarrow n +
\nu_e$, 
also known as nucleon direct Urca process (or cycle), are only
possible in neutron stars if the proton fraction exceeds a critical
threshold \cite{lattimer91:a}. 
Otherwise energy and momentum can not
be conserved simultaneously for these reactions. For a neutron star
made up of only neutrons, protons and electrons, the critical proton
fraction is around $11\%$. This follows readily from ${\mathbf
k}_{F_n} = {\mathbf k}_{F_p} + {\mathbf k}_{F_e}$ combined with the
condition of electric charge neutrality of neutron star matter. The
triangle inequality then requires for the magnitudes of the particle
momenta $k_{F_n} \leq k_{F_p} + k_{F_e}$, and charge neutrality
constrains the particle Fermi momenta according to $\kFp =
\kFe$.  Substituting $\kFp = \kFe$ into the triangle inequality leads
to $\kFn \leq 2 \kFp$ so that $n_n \leq 8 n_p$ for the number
densities of neutrons and protons.  
Expressed as a fraction of the
system's total baryon number density, $n\equiv n_p + n_n$,
one thus arrives at $n_p / n > 1/9 \simeq 0.11$ as quoted above.
Medium effects and interactions among the particles modify this value
only slightly but the presence of muons raise it to about $0.15$.  
Hyperons, which may exist in neutron star matter
rather abundantly, produce neutrinos via the direct Urca process
$\Sigma^- \rightarrow \Lambda + e^- + \bar{\nu}_e$ with $\Lambda + e^-
\rightarrow \Sigma^- + \nu_e$ and similar ones involving hyperons and nucleons
simultaneously \cite{prakash92:a,haensel94:a}.  
The cooling driven by the nucleon
direct Urca, however, dominates over the energy loss produced by the
direct Urca process among hyperons \cite{schaab95:b}.

{\bf Meson Condensate Urca Processes.} The pion or kaon meson fields
may develop condensates in dense neutron star matter.  These
condensates would have two important effects on neutron stars.
Firstly, they would soften the EOS above the critical
density for onset of condensation, which reduces the maximal possible
neutron star mass. 
Secondly, since the condensate, $<\!\!\pi^-\!\!>$ or $<\!\!K^-\!\!>$, 
can absorb as little
or as much momentum as required by the scattering processes 
$n + <\!\!\pi^-\!\!> \rightarrow n + e^- + \bar\nu_e$ or 
$n + <\!\!K^-\!\!> \rightarrow n + e^- + \bar\nu_e$,
the resulting neutrino emissivities of
meson-condensed matter \cite{MBCDM77,BKPP88,T88}, even though not as
high as the ones of the direct Urca processes (see
Table~\ref{tab:emis.core}), still lead to fast cooling
\cite{PB90,UNTTT94,schaab95:a}.
Since the $K^-$-condensate process involves strangeness change it is
less efficient that the $\pi^-$-condensate process, roughly by a factor
$\sin^2 \theta_C \simeq 1/20$ ($\theta_C$ being the Cabibbo angle).
However, medium effects can reduce the $\pi^-$-condensate process by
about one order of magnitude and make it comparable to the 
$K^-$-condensate one \cite{T83}.
Estimates predict the onset of charged pion condensation at a density
$n_\mathrm{cr}^\pi \sim 2 n_0$
($n_0=0.16~{\rm fm}^{-3}$ being the empirical nuclear matter density).  
However, this value is very sensitive to the strength of the effective 
nucleon particle-hole repulsion in the isospin-1, spin-1 channel, 
which tends to suppress the $\pi$-condensation mechanism and may push
$n_\mathrm{cr}^\pi$ to much higher values.
Similarly, depending on the nuclear model, the threshold density 
for the onset of kaon condensation, $n_\mathrm{cr}^K$, is at least
of the order of $4 n_0$ \cite{waas97:a}.

{\bf Modified Urca Processes.}  In absence of hyperons or meson
condensates, or in case the proton fraction is below threshold, none
of the above described Urca processes is possible.  In this case, the
dominant neutrino emission process is a second order process, variant
of the direct Urca process, called modified Urca process
\cite{CS64,FM79}, in which a bystander neutron or proton participates
to allow momentum conservation (see Table \ref{tab:emis.core}).  Since
this modified Urca process involves 5 degenerate fermions, instead of
three for the direct Urca and meson Urca processes, its efficiency is
reduced, simply by phase space limitation, by a factor of order
$(T/T_\mathrm{F})^2$.  This reduction, for $T_\mathrm{F} \sim 100$ MeV
and $T = 0.1$ MeV $\simeq 10^9$ K, amounts to about 6 order of
magnitude (!) and an overall temperature dependence $T^8$ instead of
$T^6$.  It is certainly the dominant process for not to high densities
in absence of pairing, and is the essence of the old ``Standard
Cooling Scenario''. However, in presence of pairing, neutrino emission by the
constant formation of Cooper pairs, see
\S~\ref{sec:pairing}, most probably dominates over the modified Urca
process.

Since the modified Urca process involves a strong interaction for the
momentum exchange between the neutrino emitting nucleon and the bystander
one, it is prone to medium corrections which seem to result in a reduction
of emissivity (see. e.g., \cite{vDDT03,CP02}).
However, softening of the pion mode, which eventually leads to
$\pi^-$-condensation, do results in a very strong enhancement of the emissivity
when the density approaches $n_\mathrm{cr}^\pi$, and gives a smooth transition
from the modified Urca process toward the $\pi^-$-condensate process through
a {\em medium-modified-Urca} process (``MMU'' process \cite{V01,VS86}).

\begin{table}[t]
\begin{center}
\caption{Dominant neutrino emitting processes in deconfined quark matter.}
\label{tab:emis.quark}
\begin{tabular}{llcc} 
\hline 
     Name           &               Process$^a$                 &       Emissivity$^b$       & Efficiency    \\ 
                    &                                        &   (erg cm$^{-3}$ s$^{-1}$) &               \\
\hline 
\parbox[c]{3.5cm}{Direct Urca cycle \\ ($ud$ branch)}         &
\rule[-0.4cm]{0.02cm}{0.85cm}
\bn u+e^- \rightarrow d+\nu_e \\ d \rightarrow u+e^-+\bar\nu_e \en
                                                                &     $\sim 10^{26} \: R \: T_9^6$    & Fast \\
\parbox[c]{3.5cm}{Direct Urca cycle \\ ($us$ branch)}         &
\rule[-0.4cm]{0.02cm}{0.85cm}
\bn u+e^- \rightarrow s+\nu_e \\ s \rightarrow u+e^-+\bar\nu_e \en
                                                                &     $\sim 10^{25} \: R \: T_9^6$    & Fast \\
\parbox[c]{3.5cm}{Modified Urca cycle \\ ($ud$ branch)}      &
\rule[-0.4cm]{0.02cm}{0.85cm}
\bn Q+u+e^- \rightarrow Q+d+\nu_e \\ Q+d \rightarrow Q+u+e^-+\bar\nu_e \en
                                                                &     $\sim 10^{21} \: R \: T_9^8$    & Slow \\
\parbox[c]{3.5cm}{Modified Urca cycle \\ ($us$ branch)}         &
\rule[-0.4cm]{0.02cm}{0.85cm}
\bn Q+u+e^- \rightarrow Q+s+\nu_e \\ Q+s \rightarrow Q+u+e^-\bar\nu_e \en
                                                                &     $\sim 10^{20} \: R \: T_9^8$    & Slow \\
Bremsstrahlungs &$Q_1+Q_2 \rightarrow Q_1+Q_2+\nu+\bar\nu$      &     $\sim 10^{19} \: R \: T_9^8$    & Slow \\
\parbox[c]{3.5cm}{Cooper pair \\ formations}          &
\bn    u+u \rightarrow [uu] +\nu+\bar\nu \\ d+d \rightarrow [dd] +\nu+\bar\nu \vspace{-0.2cm}
\\ s+s \rightarrow [ss] +\nu+\bar\nu \en  & 
\bn \sim 2.5\!\!\times\!\! 10^{20} \: R \: T_9^7 \\ \sim 1.5\!\!\times\!\! 10^{21} \: R \: T_9^7 \vspace{-0.2cm}
\\ \sim 1.5\!\!\times\!\! 10^{21} \: R \: T_9^7 \en & Slow \\
\hline 
\end{tabular}
\end{center}
\vspace*{.4cm}
\noindent
$^a$ Muons are never present in quark matter, but in case $e^+$ are present instead of
     $e^-$, processes involving $e^-$ are replaced by similar ones involving $e^+$. \\
$^b$ Quoted emissivities are only indicative: each process has its specific dependences on
     medium and particle densities, strange quark mass $m_s$, and color coupling constant $\alpha_c$,
     see Ref.~\cite{weber99:book,I82,JP01} for more details; the many $R$ factors are the respective
     control functions which take into account the effects of pairing as discussed in
     \S~\ref{sec:pairing}. 
\end{table}

{\bf The Quark Urca Processes.}
The neutrino emission processes in non-super\-conducting quark matter \cite{I82} 
can be divided into slow and fast ones, in complete analogy to the nucleon
processes discussed above, and the dominant processes are listed
in Table~\ref{tab:emis.quark}.
The fast quark direct Urca processes $d \rightarrow u + e^- + \bar{\nu}_e$ 
and $s \rightarrow u + e^- + \bar{\nu}_e$ are only possible if the Fermi 
momenta of quarks and electrons obey the triangle inequalities
associated with these reactions, which are $k_{F_d} < k_{F_u} +
k_{F_e}$ and $k_{F_s} < k_{F_u} + k_{F_e}$.  If the electron Fermi
momentum, $k_{F_e}$, in quark matter is too small for the quark
triangle inequalities to be fulfilled, a bystander quark is needed to
ensure energy and momentum conservation in the scattering
process. This latter case is referred to as quark modified Urca
process. The emissivities associated with the quark modified Urca
processes are considerably smaller than those of the direct Urca
processes because of the different phase spaces associated with
two-quark scattering and quark decay. In the extreme case when the
electron fraction vanishes entirely in quark matter, both the quark
direct and the quark modified Urca processes become unimportant.  The
neutrino emission from the star is then dominated by bremsstrahlung
processes, $Q_1 + Q_2
\rightarrow Q_1 + Q_2 + \nu + \bar\nu$, where $Q_1$ and $Q_2$ denote 
any pair of quark flavors.  
In this case, stellar cooling proceeds rather slowly.

Notice that non-Fermi liquid corrections lead to an enhancement of the
emissivities compared to the results of \cite{I82} and Table~\ref{tab:emis.quark},
which for the direct Urca is of order $(\log(mc^2/k_BT))^2$, 
with $mc^2 \sim 400$ MeV \cite{SS05}.
However this effect has not yet been incorporated in numerical calculations
and is balanced by a similar increase in the specific heat so that the
net result of non-Fermi liquid effect is not expected to be very large
\cite{SS05}.

\vspace{-0.5cm}
\subsection{Pairing} 
           \label{sec:pairing}

Pairing will unavoidably occur in a degenerate Fermi system in case
there is {\em any} attractive interaction between the particles.  In
case of the baryons, and quarks, in the neutron star interior there
are many candidates for channels of attractive interactions, and the
real question is rather what is the critical temperature $T_\mathrm{c}$ at
which pairing occurs?
Calculation of $T_\mathrm{c}$ are notoriously difficult and results often 
highly uncertain.
We refer the reader to \cite{LS01,DHJ03} for detailed reviews.
With respect to leptons, there is no obvious attractive interaction which
could lead to pairing with a $T_\mathrm{c}$ of significant value.

\begin{figure}[tb]
\begin{center}
\includegraphics[scale=0.90]{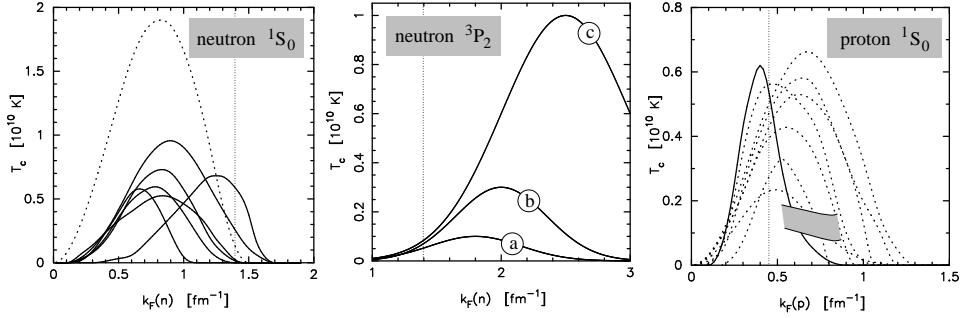}
\caption{Predictions for the nucleon pairing $T_\mathrm{c}$.
         Left panel: neutron ${^1S_0}$ pairing; all calculations include
         medium polarization in some measure except for the dotted
         curve which illustrates the strong reduction effect of
         polarization on this gap.  Right panel: same for protons; no
         polarization effects are taken into account except for the
         continuous curve; the shaded area shows the range of expected
         $T_\mathrm{c}$'s when polarization is taken into account, according to
         the estimates of \cite{AWP91}.  Central panel: neutron
         ${^3P_2}$ pairing, three typical results illustrating the
         possible range according to \cite{BEEHS98}.  See
         \cite{PLPS04} for details and references.}
\label{fig:Tc}
\end{center}
\end{figure}

In the case of nucleons, at low Fermi momenta, pairing is predicted to
occur in the ${^1S_0}$ angular momentum state while at larger momenta
neutrons are possibly paired in the ${^3P_2} - {{^3F_2}}$ state (the mixing
being due to the tensor interaction).    In the case
of the neutron ${^1S_0}$ pairing, which occurs at densities
corresponding to the crust and, possibly, the outermost part of the
core, much efforts have been invested in its study and calculations
are converging with time when more and more sophisticated many-body
models are used.  In the case of the proton ${^1S_0}$ pairing the
situation is more delicate since it occurs at densities (in the outer
core) where protons are mixed, to a small amount, with
neutrons. Predictions for $T_\mathrm{c}$ span a much wider range than in the
case of the neutron ${^1S_0}$ gap. Whether or not neutrons pair in the
${^3P_2} - {^3F_2}$ channel is still uncertain, since even the best
available models for the nucleon-nucleon interaction fail to reproduce
the measured ${^3P_2}$ phase shift {\em in vacuum} \cite{BEEHS98}.
Moreover, the results of \cite{SF04}, which consider
polarization contributions to the effective interaction, indicate
that this gap may be vanishingly small. A set of representatives
predictions for the nucleon gaps are shown in Figure~\ref{fig:Tc}.

\begin{figure}[tb]
\begin{center}
\includegraphics[scale=0.50]{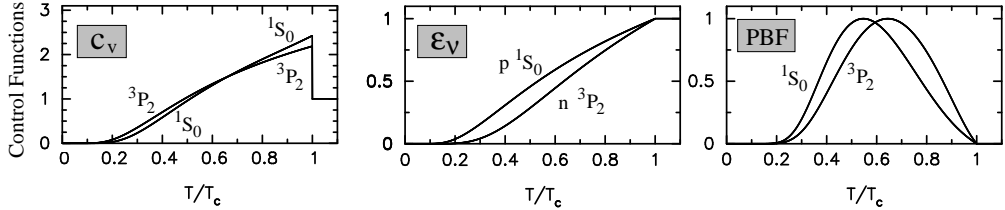}
\caption{Control functions for $c_\mathrm{v}$ (left panel), 
Eq.~(\protect\ref{equ:cv_paired}), $\epsilon_\nu$ of the modified Urca process
(central panel), Eq.~(\ref{equ:nu_paired}), and the PBF process, 
Eq.~(\protect\ref{Eq:PBF2})
(right panel).}
\label{fig:control}
\end{center}
\end{figure}

The enormous impact of pairing on the cooling comes directly from the
appearance of the energy gap $\Delta$ at the Fermi surface which leads
to a suppression of all processes involving single particle
excitations of the paired species.  When $T \ll T_\mathrm{c}$ the suppression
is of the order of $e^{-\Delta/k_\mathrm{B} T}$ and hence dramatic.  The
suppression depends on the temperature dependence of $\Delta$ and the
details of the phase space involved in each specific process. 
In numerical calculations it is introduced as a control function. 
For the specific heat one has
\be
c_\mathrm{v}(T) \longrightarrow c_\mathrm{v}^\mathrm{paired}(T) = R_c(T/T_\mathrm{c}) \times
c_\mathrm{v}^\mathrm{normal}(T) \, ,
\label{equ:cv_paired}
\ee
and the control functions have been calculated for both ${^1S_0}$ and
${^3P_2}$ pairing in \cite{LY94a}.  For neutrino processes there is a
long family of control functions for all processes which must also
consider which of the participating baryons are paired.  
As for $c_\mathrm{v}$ one uses
\be
\epsilon_\nu(T) \longrightarrow 
    \epsilon_\nu^\mathrm{paired}(T) = R_\nu(T/T_\mathrm{c}) \times
    \epsilon_\nu^\mathrm{normal}(T) \, ,
\label{equ:nu_paired}
\ee
and the $R_\nu$'s for many processes can be found in \cite{YKGH01}.
We plot in Figure~\ref{fig:control} several examples of control functions.

It is important to notice that the gap $\Delta$ is actually a function of the
particle momentum, $\Delta = \Delta(\mathbf{k})$. 
For ${^1S_0}$ pairing it is isotropic, i.e., $\Delta = \Delta(k)$, but for
${^3P_2} - {^3F_2}$ pairing the angular dependence of $\Delta(\mathbf{k})$ is
complicated.
In this latter case many phases, with distinct angular dependences of
$\Delta(\mathbf{k})$ are possible (\cite{ZCK03} found there are {\em at least} 13 of them)
and for several of them $\Delta(\mathbf{k})$ has nodes at some points of along
some lines on the Fermi surface. 
The control functions plotted in Figure~\ref{fig:control} assume nodeless gaps,
but in cases of 1D nodes $R \sim (T/T_c)^2$, while for 2D nodes $R \sim T/T_c$,
at $T \ll T_c$ instead of a Boltzmann-like suppression.

{\bf Cooper Pair Breaking and Formation (PBF) Processes}.
Besides the above described, and well known, suppressing effects on
the specific heat and neutrino emissivities, the onset of pairing also
opens new channels for neutrino emission.  The superfluid or
superconducting condensate is in thermal equilibrium with the single
particle (``broken pairs'') excitations and there is continuous formation 
and breaking of Cooper pairs, which are very intense at temperatures 
slightly below $T_\mathrm{c}$.  
The formation of a Cooper pair liberates an energy
which can be taken away by a $\nu$-$\overline{\nu}$ pair
\cite{FRS76,VS87}
\be
X + X \longrightarrow [XX] + \nu + \overline{\nu} 
\label{Eq:PBF}
\ee
where $[XX]$ denotes a Cooper pair of particles $X$ ($X$ stands for
neutrons, protons, hyperons, quarks, etc.).  As an example, the
emissivity for neutron ${^3P_2}$ pairing is \cite{YKGH01}
\footnote{Our control function $\tilde{R}$ differs from the $R$ in
\protect\cite{YKGH01} in that it is normalized to have a maximum value of one.} 
\be
q_\nu^{n,{}  \; {^3P_2}} =
   8.6 \times 10^{21} \left(\frac{n_\mathrm{b}}{n_0}\right)^{1/3}
   \left(\frac{m_n^*}{m_n}\right)  \times
   \nonumber \\
   \tilde{R}_{{^3P_2}}(T/T_\mathrm{c}) 
   \left(\frac{T}{10^9 ~\mathrm{K}}\right)^7 \, ,
\label{Eq:PBF2}
\ee
where the control function $\tilde{R}$ is plotted in the right
panel of Figure~\ref{fig:control}: the process turns on at $T =
T_\mathrm{c}$, with an increasing efficiency when $T$ decreases, since
the energy of the emitted neutrinos is determined by the gap's size
which grows with decreasing temperature just below $T_\mathrm{c}$, and
is eventually exponentially suppressed when $T \ll T_\mathrm{c}$ as
pair breaking is frozen because $k_\mathrm{B} T \ll \Delta$.  
This process can be seen as a bremsstrahlung with a very strong correlation
in the final state and
referring to Tables~\ref{tab:emis.core} and \ref{tab:emis.quark}, one sees
that it is much more efficient that the simple bremsstrahlung one 
and it can even dominate over the standard modified Urca process.  
This process, which is as standard as the modified Urca process, is an essential
ingredient of the Minimal Cooling Paradigm described in \S~\ref{sec:minimal}.  
Analogous processes occur for all cases of pairing: neutron, proton, hyperons,
and quarks \cite{JP01}.

\subsection{Color-Superconductivity of Quark Matter}
           \label{sec:CSC}

Already several decades ago it had been suggested that the attractive
force among quarks may cause them to form Cooper pairs
\cite{bailin84:a,bailin79:a}.  Originally the gap was estimated to be
around $\Delta = 0.4$~MeV \cite{bailin79:a}. Recently, however, it was
discovered that the condensation patterns of quark matter are much
more complex than originally thought
\cite{rajagopal01:a,alford01:a}. This has its origin in the fact that
quarks come in different colors, different flavors, and different
masses. Moreover, bulk matter is neutral with respect to both electric
and color charge, and is in chemical equilibration under the weak
interaction processes that turn one quark flavor into
another. Depending on density and temperature, quarks may thus
condense in one of the following pairing schemes.  At asymptotic
densities the ground state of (3-flavor) QCD with a vanishing strange
quark mass is the color-flavor locked (CFL) phase in which all three
quark flavors participate symmetrically. This phase has been shown to
be electrically neutral without any need for electrons for a
significant range of chemical potentials and strange quark masses
\cite{rajagopal01:b}.  In the opposite limit where the strange quark
mass $m_s$ is large enough that strange quarks can be ignored, then up
and down quarks may pair in the 2-flavor superconducting (2SC) phase.
Other possible condensation patterns are CFL-$K^0$, CFL-$K^+$ and
CFL-$\pi^{0,-}$, gCFL (gapless CFL phase), 1SC
(single-flavor-pairing), CSL (color-spin locked phase), and the LOFF
(Larkin, Ovchinnikov, Fulde, and Ferrell) crystalline pairing phase,
depending on the quark flavor densities, the quark chemical potential
$\mu_q$, and electric charge density.  For chemical potentials that
are of astrophysical interest, $\mu_q < 1000$~MeV, the color gap is
between 50 and 100~MeV, which has a significant impact on stellar
cooling by neutrino emission.

\vspace{-0.5cm}
\subsection{The Surface Photon Luminosity and the Envelope} 
           \label{sec:envelope}
\vspace{-0.5cm}

The photon luminosity $L_\gamma$ is traditionally expressed as
\be
L_\gamma = 4 \pi R^2 \; \cdot \; \sigma_\mathrm{\scriptscriptstyle SB}
T_\mathrm{e}^4 \, ,
\label{equ:L}
\ee
which {\em defines} the effective temperature $T_\mathrm{e}$
($\sigma_\mathrm{\scriptscriptstyle SB}$ being the Stefan-Boltzmann
constant and $R$ the stellar radius). 
The quantities $L$, $R$, and $T_\mathrm{e}$ are local quantities
as measured by an observer at the stellar surface.  An external
observer ``at infinity'' will measure these quantities red-shifted,
i.e., $L_\infty = e^{2\phi} L_\gamma$, $T_\infty = e^\phi T_\mathrm{e}$,
and $R_\infty = e^{-\phi} R$, 
where $e^{2\phi} \equiv g_{00}$ is the time component of the metric and is
related to the red-shift $z$ by $e^{-\phi} = 1+z$, so that
\be
L_\infty = 4 \pi R_\infty^2 \; \cdot \; \sigma_\mathrm{\scriptscriptstyle SB}
T_\infty^4 \, .
\label{equ:Linf}
\ee
The luminosity $L_\gamma$, or $L_\infty$, is the main output of a cooling
calculation, and it can equally well be expressed in terms of
$T_\mathrm{e}$ or $T_\infty$.  

Numerical simulations calculate the time evolution of the internal temperature 
$T = T(\rho,t)$ and luminosity $L = L(\rho,t)$ profiles (viewed as functions 
of the density $\rho$ instead of the radius $r$) up to an outer boundary 
$\rho_\mathrm{b}$.
This boundary is chosen such that at this point the diffusive luminosity
$L(\rho_\mathrm{b})$ is equal to the photon luminosity at the surface,
i.e., $L(\rho_\mathrm{b}) \equiv L_\gamma$, and an envelope model
is glued as an outer boundary condition.
Typically $\rho_\mathrm{b}$ is taken as $10^{10}$~g~cm$^{-3}$ and the
envelope is thus a thin layer, of the order of a hundred meters thick,
which is treated in the plane parallel approximation.  Assuming that
the thermal relaxation time-scale of the envelope is much shorter than
the stellar evolution time-scale, and that neutrino emission in the
envelope is negligible, hydrostatic equilibrium and heat transport
reduce to ordinary differential equations which, with the appropriate
physical input, are easily solved.  The result is a surface
temperature $T_\mathrm{s} \equiv T_\mathrm{e}$ for each given
$T_\mathrm{b} \equiv T(\rho_\mathrm{b})$.  It is usually called a
$T_\mathrm{b} - T_\mathrm{s}$ or $T_\mathrm{b} - T_\mathrm{e}$
relationship.  Through Eq.~(\ref{equ:L}), this gives us a relationship
between $L(\rho_\mathrm{b}) \equiv L_\gamma$ and $T(\rho_\mathrm{b})$
which is the outer boundary condition for the cooling code.

It has been shown in Refs.\ \cite{GPE82,GPE83} that $T_\mathrm{e}$ is
actually controlled by a ``sensitivity layer'', where electrons
become partially degenerate and ions are in the liquid phase. At
higher densities the highly degenerate electrons are extremely
efficient in transporting heat while at lower densities photons take
over and are also very efficient. 
The density at which the sensitivity layer
is located increases with increasing $T_\mathrm{b}$.
This sensitivity layer is hence a throttle and once heat has passed
through it it can freely flow to the surface and be radiated.
The layers at densities below the sensitivity layer have no effect at all 
on the thermal evolution of the star, since they are unable to alter the 
heat flow, but the outermost layer, the photosphere, is of course of upmost
observational importance since it is there that the energy distribution
of the emerging flux, i.e., the observable spectrum, is determined.

\begin{figure}[t]
\begin{center}
\includegraphics[scale=0.6]{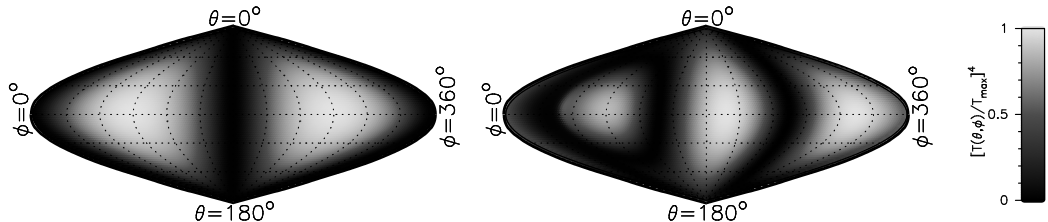}
\caption{Two examples of surface temperature distributions induced by
the magnetic field, in an area preserving projection of the neutron
star surface (grey shading, shown on the right scale, follows the
surface flux instead of the temperature).  Left panel assumes a
dipolar field, with strength $1.2\times 10^{12}$ G at the pole
located at $(\theta,\phi) = (90^o,90^o)$: for a core temperature of
$4.05\times10^7$ K it gives $T_\mathrm{e} = 5.43\times10^5$ K (see
Eq.~\protect\ref{Eq:Teff-B}) while the maximum and minimal surface
temperatures, at the magnetic poles and along the magnetic equator,
respectively, are $T_\mathrm{max} = 6.70\times10^5$ K and
$T_\mathrm{min} = 1.4\times10^5$ K.  The right panels shows the effect
the same dipolar field to which a quadrupolar component has been
added: this results in $T_\mathrm{e} = 5.31\times10^5$ K.  This
particular latter case allows to reproduce the observed ROSAT X-ray
pulse profile of Geminga (see Figure~6 in \cite{PS96}) which shows a
single very broad pulse while a purely dipolar field would result in a
double pulse profile (assuming the observer is in the direction
$\theta \simeq 90^o$ and emission is isotropic blackbody).  Finally,
in absence of magnetic field, the same internal temperature would
result in $T_\mathrm{e} = 5.54\times10^5$ K.  }
\label{fig:Tdistr1}
\end{center}
\end{figure}

Gluing an envelope to an interior solution is a standard technique in
stellar evolution codes.  For neutron stars it has two extra
advantages: it relieves us from solving for hydrostatic equilibrium in
the interior, since matter there is degenerate, and, most importantly,
it allows one to easily include magnetic field effects.
The magnetic field slightly enhances heat transport
along it but strongly suppresses it in the perpendicular direction,
resulting in a highly non uniform surface temperature \cite{GH83}.
Assuming that magnetic field effects on heat transport
are negligible at $\rho > \rho_\mathrm{b}$ one keeps spherical
symmetry in the interior and thus has a unique $T_\mathrm{b}$ at
$\rho_\mathrm{b}$.  
For this given uniform $T_\mathrm{b}$ one can 
piece together a set of envelope calculations for the various field 
strengths and orientations along the stellar surface,
corresponding to the assumed magnetic field structure, and thus obtain
a non uniform surface temperature distribution $T_s(\theta,\phi)$,
in spherical coordinates $(\theta,\phi)$ \cite{P95,PS96}.
The effective temperature is simply obtained by averaging the
locally emerging photon flux 
$F_\gamma(\theta,\phi) \equiv \sigma_\mathrm{SB} \, T_\mathrm{s}^4(\theta,\phi)$
over the whole stellar surface
\be
T_\mathrm{e}^4 \equiv 
\frac{1}{4\pi} \int \!\! \int T_\mathrm{s}^4(\theta,\phi) \; \sin \theta \, d\theta d\phi
\label{Eq:Teff-B}
\ee
Two examples of such temperature distributions are illustrated in Figure~\ref{fig:Tdistr1}.
The overall effect on $T_\mathrm{e}$ is nevertheless surprisingly small,
see, e.g., \cite{P95,PS96} and the examples in Figure~\ref{fig:Tdistr1},
and a non-magnetic envelope is actually a rather good approximation.
However, the assumption of spherical symmetry at $\rho >
\rho_\mathrm{b}$ is questionable and will be discussed in
\S~\ref{sec:magcrust}.

\begin{figure}[t]
\begin{center}
\includegraphics[scale=0.3]{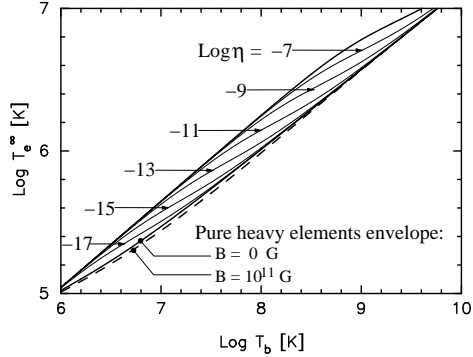}
\caption{Relationship between the red-shifted effective temperature
$T_\mathrm{e}^\infty$ and the interior temperature $T_\mathrm{b}$ at the bottom of
the envelope assuming various amounts of light elements parameterized
by $\eta \equiv g_{\mathrm{s} \, 14}^2 \Delta M_\mathrm{L}/M$ ($\Delta
M_\mathrm{L}$ is the mass in light elements in the envelope,
$g_{\mathrm{s} \, 14}$ the surface gravity in units of $10^{14}$ cm
s$^{-2}$, and $M$ is the star's mass), in the absence of a magnetic
field \cite{PCY97}.  Also shown are the $T_\mathrm{b} -
T_\mathrm{e}^\infty$ relationships for an envelope of heavy elements
with and without the presence of a dipolar field of strength of
$10^{11}$ G following \cite{PY01}.  Notice that the smaller is $\Delta
M_\mathrm{L}$ the lower is the temperature at which its effect is
felt.}
\label{fig:envelope}
\end{center}
\end{figure}

Given that the overall effect of the magnetic field, in the envelope, is not
very strong, it turns out that the major uncertainty about the envelope
is its chemical composition.  The standard neutron star crust is made of
cold catalyzed matter, which means $^{56}$Fe at low density 
($\rho < 10^6$ g cm$^{-3}$).  
However real neutron stars may be dirty and have lighter elements at their
surface.  As was shown in \cite{CPY97} the presence of light elements
in the envelope strongly enhances heat transport (e.g., the electron
thermal conductivity within liquid ions of charge $Z$, in the
sensitivity layer, is roughly
proportional to 1/$Z$, \cite{YU80}) and results in a
significantly higher $T_\mathrm{e}$, for the same $T_\mathrm{b}$, than
in the case of a heavy element envelope.  
Due to pycnonuclear fusion,
light elements are unlikely to be present at densities above
$10^9$~g~cm$^{-3}$. At very high $T_\mathrm{e}$ this density is below
the sensitivity layer and light elements have little effect, but for
$T_\mathrm{e}$ within the observed range ($\sim 10^5 - 10^6$ K) the
sensitivity layer is at a sufficiently low density so that it can easily 
be contaminated with light elements and the $T_\mathrm{b} -
T_\mathrm{e}$ relationship can be significantly altered. However, if only a
small amount of light elements is present at the surface their effect
will only be felt a low $T_\mathrm{e}$.  We show in
Figure~\ref{fig:envelope} the $T_\mathrm{b} - T_\mathrm{e}$
relationships for various amounts of light elements and also, for
comparison, the case of a magnetized envelope with a $10^{11}$ G
dipolar magnetic field.

\subsection{Comments on minor neutrino emission processes}
           \label{sec:minor-nu}

The neutrino processes described in \S~\ref{sec:neutrinos} are usually the
dominant ones but in particular cases some minor processes may become important.

During the first few years of the life of the compact star the surface 
temperature is entirely controlled by neutrino emission in its upper layers and
is independent of what is happening in the core \cite{BKPP88,NT87,LvRPP94,GYP01}.
In this case the dominant process is plasmon decay (see, e.g., \cite{YKGH01})
which very efficiently cools every layer in the the outer crust until the temperature 
$T$ drops below $\hbar \Omega_P/k_B$ 
($\Omega_P$ being the electron plasma frequency in the layer)
and plasmons are exponentially suppressed. 
This leads to some almost universal temperature $T_e \simeq 2.5 \times 10^6$ K
(in case of a heavy elements envelope) for all young compact stars.
Later on the crust will cool by neutrino emission from the
electron-ion and electron-electron bremsstrahlung processes
\cite{YKGH01,JGP05} and the PBF process from the neutron $^1$S$_0$ superfluid
in the inner crust.
This is not very important at present time since there are no data
about such young compact stars, but may become essential in case a compact object
is detected within the remnant of the supernova SN 1987A.

For middle-age stars, in case all processes described in \S~\ref{sec:neutrinos}
and \S~\ref{sec:pairing} are strongly  suppressed by very large gaps then
neutrino emission from the crust, as described in the previous paragraph, is
important.
Despite of the small amount of mass present in the crust, omission of its 
neutrino emission may then lead to erroneous results, i.e., too warm stars.
All numerical results presented in this paper of course include these processes.

\section{Some simple analytical solutions}
        \label{sec:analytical}

Assuming simplified physics input one can easily obtain some very
illustrative analytical solutions to the cooling
Eq.~(\ref{equ:energy-conservation}).  Let us write
\be
C_\mathrm{v} = C \cdot T \, ,
\;\;\;\;\;\;\;\;\;\;
L_\nu^\mathrm{slow} = N^\mathrm{s} \cdot T^8 \, ,
\;\;\;\;\;\;\;\;\;\;
L_\nu^\mathrm{fast} = N^\mathrm{f} \cdot T^6 \, ,
\label{equ:aprox1}
\ee
for the specific heat from degenerate fermions and neutrino emission
with only ``slow'' processes or with some of the ``fast'' processes
listed in Table~\ref{tab:emis.core}.  For $L_\gamma$ we write
\be
L_\gamma \equiv 4 \pi R^2 \sigma_\mathrm{\scriptscriptstyle SB} T_\mathrm{e}^4 = S
T^{2+4\alpha} 
\;\;\;\;\;\;
\mathrm{using}
\;\;\;\;\;\;
T_\mathrm{e} \propto T^{0.5 + \alpha} 
\;\;\;\;
(\alpha \ll 1) \, ,
\label{equ:aprox2}
\ee
where $T_\mathrm{e}$ has been converted into the internal temperature $T$ through
an envelope model with a power-law dependence.
Figure~\ref{fig:envelope} shows that for a heavy element envelope or
an envelope with a large amount of light elements this power-law
relation is a good approximation in a wide range of temperatures.
Typical value for the numerical coefficients in Eqs.~(\ref{equ:aprox1})
and (\ref{equ:aprox2}) are listed in Table~\ref{tab:aprox}.

\begin{table}[tb]
\begin{center}
\caption{Typical numerical coefficients for simplified power-law cooling 
models}
\label{tab:aprox}
\begin{tabular}{lccccc} 
\hline
       & $C$ & $N^\mathrm{s}$ & $N^\mathrm{f}$ & $S$ & $\alpha$ \\ & erg
       K$^{-2}$ & erg s$^{-1}$ K$^{-8}$ & erg s$^{-1}$ K$^{-6}$ & erg
       s$^{-1}$ K$^{-2-4\alpha}$ & \\

\hline
high   &  $10^{30}$   &     $10^{-32}$        &     $10^{-9}$         &      $4 \times 10^{14}$     & 0.1  \\
low    &  $10^{29}$   &         0             &         0             &      $2 \times 10^{15}$     & 0.05 \\
\hline
\end{tabular}
\end{center}
\begin{scriptsize}
Comments:\\ 
$C$: ``low'' value corresponds to the leptons contribution
only, i.e., assuming all baryons are paired with high
$T_\mathrm{c}$'s, while ``high'' value corresponds to total absence of
pairing. \\ 
$N^\mathrm{s}$: ``low'' value corresponds to complete
baryon pairing and ``high'' value to total absence of pairing. \\
$N^\mathrm{f}$: ``high'' value corresponds approximately to an inner
core of 1 \Msun sustaining a direct Urca process with nucleons,
``low'' value of course corresponds to total absence of fast neutrino
emission. \\ 
In case of total baryon pairing $N^\mathrm{s} \simeq
N^\mathrm{f} \simeq 0$: in this case one has to consider neutrino
emission in the crystallized crust by the e-ion bremsstrahlung process with a
luminosity of the order of $L_\nu^\mathrm{cr} = N^\mathrm{cr} \cdot T^8$ 
and $N^\mathrm{cr} \sim 10^{-34}$ erg s$^{-1}$ K$^{-8}$. \\ 
S
and $\alpha$: ``high'' corresponds to an envelope with a maximum
amount of light elements while ``low'' corresponds to an envelope made
of heavy elements, in both cases with a stellar radius $R \sim 12$ km.
\end{scriptsize} 
\vspace{+0.5cm}
\end{table}

Due to the much stronger $T$ dependence of $L_\nu$ compared to
$L_\gamma$, at early times neutrino emission drives the cooling
and when $T$ has sufficiently decreased photons will take over.

\noindent (1) During the {\bf neutrino cooling era} we can neglect 
$L_\gamma$ in Eq.~(\ref{equ:energy-conservation}) and, with our
approximate formulas of Eq.~(\ref{equ:aprox1}), obtain analytical
solutions
\be
\!\!\!\!\!\!\!\!\!\!
\left. \begin{array}{l} \mathrm{Slow} \; \nu \\ \mathrm{cooling} \end{array} \right\}\!\! : \;\;
t = \frac{C}{6 N^\mathrm{s}} \left( \frac{1}{T^6}-\frac{1}{T_0^6} \right)
\;\;\;
\left. \begin{array}{l} \mathrm{Fast} \; \nu \\ \mathrm{cooling} \end{array} \right\}\!\! : \;\;
t = \frac{C}{4 N^\mathrm{f}} \left( \frac{1}{T^4}-\frac{1}{T_0^4} \right)  ,
\label{Eq:neutrino-cooling1}
\ee
where $T_0$ is the initial temperature at time $t_0 \equiv 0$.  For $T
\ll T_0$, this gives, for slow $\nu$ cooling
\be
T = \left(\frac{C}{6N^\mathrm{s}}\right)^{\frac{1}{6}} t^{-\frac{1}{6}} 
\;\;\;\;\;\; \mathrm{and} \;\;\;\;\;\;\;
T_\mathrm{e} \appropto t^{-\frac{1}{12}} \, ,
\label{Eq:neutrino-cooling2}
\ee
and for fast $\nu$ cooling
\be
T = \left(\frac{C}{4N^\mathrm{f}}\right)^{\frac{1}{4}} t^{-\frac{1}{4}} 
\;\;\;\;\;\; \mathrm{and} \;\;\;\;\;\;\;
T_\mathrm{e} \appropto t^{-\frac{1}{8}} \,.
\label{Eq:neutrino-cooling3}
\ee
(we have used that $\alpha \sim 0$).  The very small exponent in the
$T_\mathrm{e}$ evolution during neutrino cooling is a direct consequence of the
strong temperature dependence of $L_\nu$.  The neutrino cooling time
scales are also very suggestive:
\be
\tau_\nu^\mathrm{slow} = \frac{C}{6 N^\mathrm{s} T^6}
\simeq 6 \; \mathrm{months} \cdot \left[\frac{C_{30}}{6 N^\mathrm{s}_{-32}
T_9^6}\right] \, ,
\label{Eq:neutrino-cooling4}
\ee
and 
\be
\tau_\nu^\mathrm{fast} = \frac{C}{4 N^\mathrm{f} T^4}
\simeq 4 \; \mathrm{minutes} \cdot \left[\frac{C_{30}}{6 N^\mathrm{f}_{-9}
T_9^4}\right] \, ,
\label{Eq:neutrino-cooling5}
\ee
and justify the names of ``slow'' and ``fast'' neutrino cooling!
Notice that $10^9$ K is a typical value for the baryon pairing $T_\mathrm{c}$,
and hence, in case of fast neutrino cooling, one can expect that a few
minutes after the star is born its core may become
superfluid/superconducting, and the neutrino emission very strongly
suppressed.

\noindent (2) During the {\bf photon cooling era} $(L_\gamma \gg L_\nu$) one
similarly obtains
\be
t = t_1 + \frac{C}{4 \alpha \, S} \left( \frac{1}{T^{4\alpha}} -
\frac{1}{T_1^{4\alpha}} \right) \, ,
\label{Eq:photon-cooling1}
\ee
where $T_1$ is the temperature at time $t_1$.  When $t \gg t_1$ and $T
\ll T_1$, we have
\be
T = \left(\frac{C}{4 \alpha S}\right)^{\frac{1}{4\alpha}}
t^{-\frac{1}{4\alpha}}
\;\;\;\; \mathrm{and} \;\;\;\; 
T_\mathrm{e} \appropto t^{-\frac{1}{8\alpha}} \,.
\label{Eq:photon-cooling2}
\ee
Since $\alpha \ll 1$, we see that, during the photon cooling era, the
evolution is very sensitive to the nature of the envelope, i.e.,
$\alpha$ and $S$, and to changes in the specific heat, as induced by
pairing.  Notice that in case $\alpha = 0$ one would obtain an
exponential solution instead of a power law.

\noindent The {\bf shift from neutrino to photon cooling:} the temperature
$T_\mathrm{shift}$ at which this happens is also easily estimated by
equating $L_\nu$ to $L_\gamma$, giving for slow $\nu$ cooling
\be
T_\mathrm{shift}^\mathrm{s} \simeq \left(\frac{S}{N^\mathrm{s}}\right)^{1/6}
  \sim 10^8 \; \mathrm{K}
\;\;\;\;\; \mathrm{and} \;\;\;\;\; T_e \sim 10^6 \; \mathrm{K}
\ee
and for fast $\nu$ cooling
\be
T_\mathrm{shift}^\mathrm{f} \simeq \left(\frac{S}{N^\mathrm{f}}\right)^{1/4}
  \sim 10^6 \; \mathrm{K}
\;\;\;\;\; \mathrm{and} \;\;\;\;\; T_e \sim 10^5 \; \mathrm{K}
\ee
However, for fast $\nu$ cooling $N^\mathrm{f}$ can be significantly reduced by 
pairing and hence $T_\mathrm{shift}^\mathrm{f}$ increased.
One may also want to obtain an estimate of the age $t_\mathrm{shift}$ at which this 
happens by using Eq.~\ref{Eq:neutrino-cooling2} and \ref{Eq:neutrino-cooling3}.
It is however now important to keep the $\alpha$ dependence in $L_\gamma$, Eq.~\ref{equ:aprox2},
and, for small $\alpha$ one obtains, for slow $\nu$ cooling,
$t_\mathrm{shift}^\mathrm{s} = C/6S (N^\mathrm{s}/S)^{(2/3 \; \alpha)}$ and a similar 
expression for fast $\nu$ cooling.
Given that $N/S$ is a {\em very} small number, the value of $t_\mathrm{shift}$ is
extremely sensitive to the exact value of $\alpha$, and also of $N$ which can be altered
by several orders of magnitude by pairing.
Considering moreover that these simple solutions are based on approximate formulas, 
such results for $t_\mathrm{shift}$ are not very useful in themselves
for numerical estimates but show that $t_\mathrm{shift}$ can change significantly
by small changes in the interior (through $N$) and/or surface (through $\alpha$) physics.
The numerical results presented in the following sections show that $t_\mathrm{shift}$
can vary from about $10^4$ yrs up to about $10^6$ years, for both slow and fast $\nu$ cooling.

\section{The ``Minimal Cooling'' paradigm} 
        \label{sec:minimal}

The Minimal Cooling paradigm assumes that no enhanced neutrino
emission is allowed, in particular that no form of ``exotic'' matter
is present in the inner core of the star.  It is hence a benchmark
against which observations of cooling neutron stars have to be
compared, and a discrepancy between the minimal cooling theoretical
predictions and data should be considered as strong evidence for
physics beyond minimal.  As such the minimal paradigm excludes {\em a
priori} the presence of charged meson condensates, hyperons, and/or
deconfined quark matter, but it also assumes that the growth of the
symmetry energy with density is slow enough that the nucleon direct
Urca process is forbidden.  It can be seen as a modern version of the
``Standard Cooling'' scenario but with the important difference that
the effects of nucleon pairing are wholly taken into account,
particularly the strong neutrino emission from the formation of
Cooper pairs during the pairing phase transition\footnote{The motivation 
for the renaming from ``Standard''
to ``Minimal'' is precisely that ``Standard Cooling'' is commonly
understood as ``Modified Urca Cooling'' while, in presence of pairing,
the dominant process is the PBF process.}.

\begin{figure}[t]
\begin{center}
\includegraphics[scale=0.55]{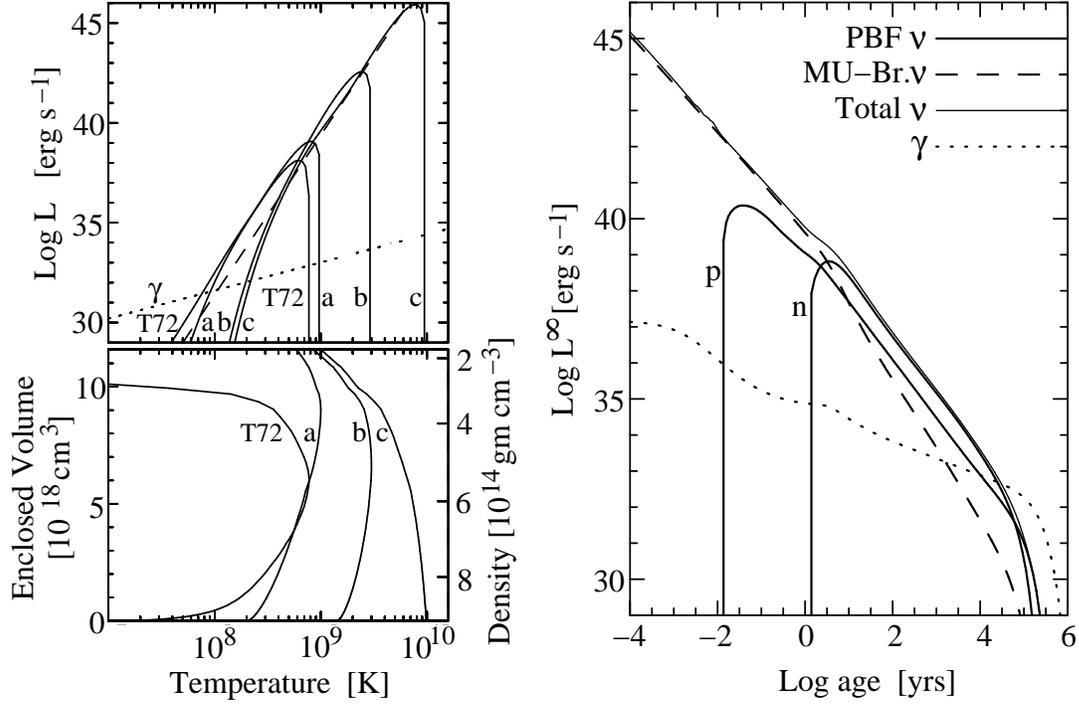}
\caption{Left Panel: neutrino luminosities from the PBF process for
         four different neutron ${^3P_2} - {^3F_2}$ gaps (labeled ``a'', ``b'', 
         and ``c' as in Figure~\ref{fig:Tc} while the gap labeled ``T72''
         is taken from \cite{T72b}).  The lower left panel shows the four
         gaps' $T_\mathrm{c}$ vs. the enclosed volume (left scale) and density
         (right scale) and the
         upper left panel the corresponding PBF luminosities as well
         as the surface photon luminosity while the dashed curve shows, for
         comparison, the total luminosity from modified Urca and nucleon
         bremsstrahlung processes {\em without pairing suppression}.
         Luminosities are calculated assuming an
         isothermal star so that,
         e.g., with a temperature of $2\times 10^9$ K, one can see that in
         case of gap ``b'', two regions have unpaired neutrons:
         the inner core at $\rho$ above 
         $\sim 8.5\times 10^{14}$ g cm$^{-3}$ and a corresponding volume of
         $\sim 1.1 \times 10^{18}$ cm$^3$, and
         the outer core at $\rho$ below $\sim 3 \times 10^{14}$ g cm$^{-3}$
         and corresponding volume of about 
         $\sim (11.5 -9.8) \times 10^{18} \simeq 1.7 \times 10^{18}$ cm$^3$,
         while the intermediate region, of volume $\simeq 9 \times 10^{18}$ cm$^3$
         has $T$ slightly below $T_c$ and produces the large $L_\nu$ by the PBF
         process shown in the upper panel.
         Right panel: comparison of luminosities from various
         processes during a realistic cooling history: photon
         (``$\gamma$''), all $\nu$-processes (``Total $\nu$''),
         modified Urca and nucleon bremsstrahlung
         (``MU-Br. $\nu$''), and PBF (``PBF $\nu$'') from $n$ ${^3P_2} -
         {^3F_2}$ and $p$ ${^1S_0}$ pairing marked by ``$n$'' and ``$p$'',
         respectively.  The $n$ ${^3P_2} - {^3F_2}$ gap is our model ``a''
         which, as shown on the left panel, is amongst the most
         efficient one for the PBF process while the $p$ ${^1S_0}$ gap
         is from \cite {AO85}.  
         All results are from Ref.~\cite{PLPS04}.}
\label{fig:lum_nu}
\end{center}
\end{figure}

\begin{figure}[t]
\includegraphics[scale=0.55]{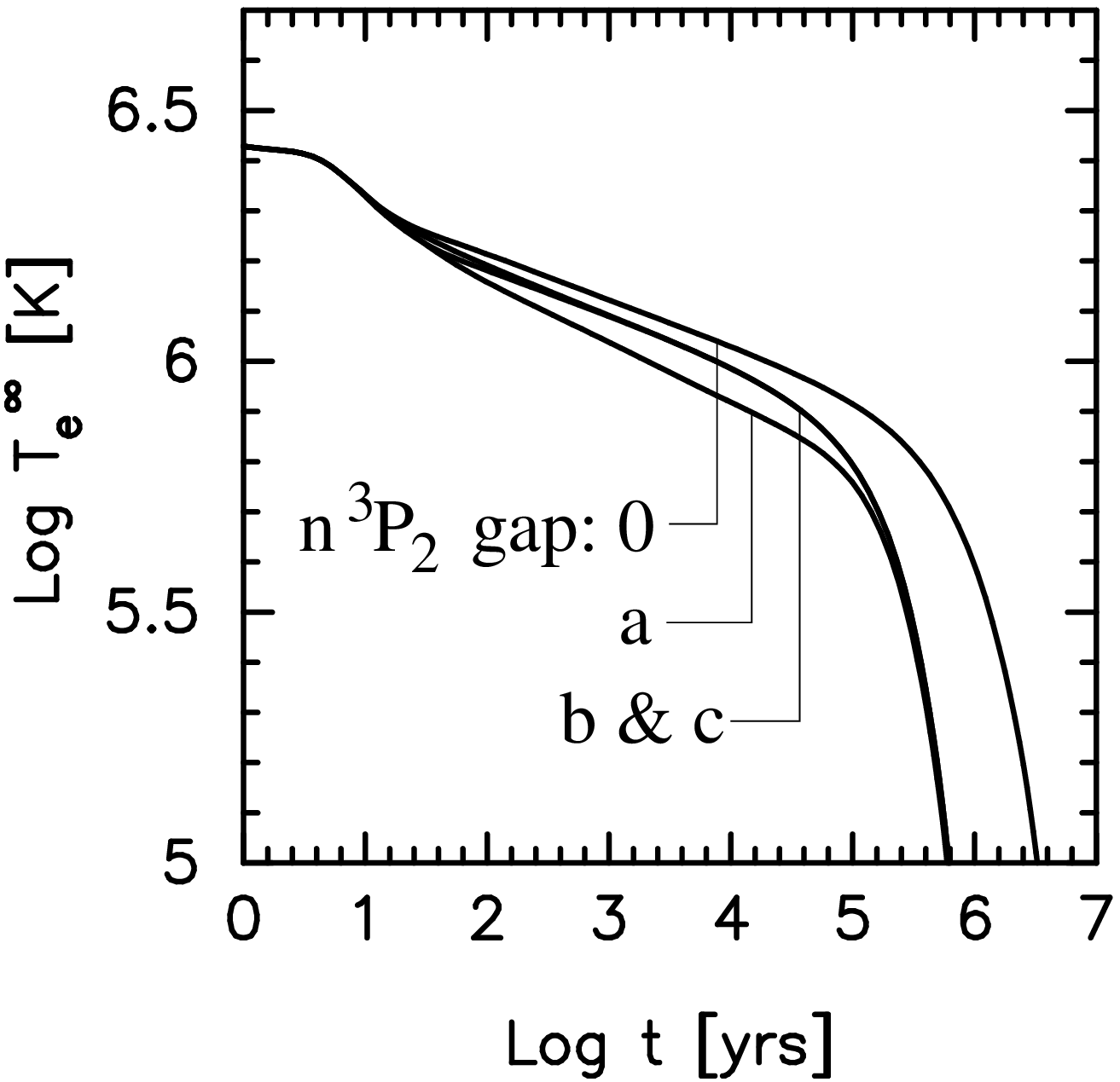}
\includegraphics[scale=0.55]{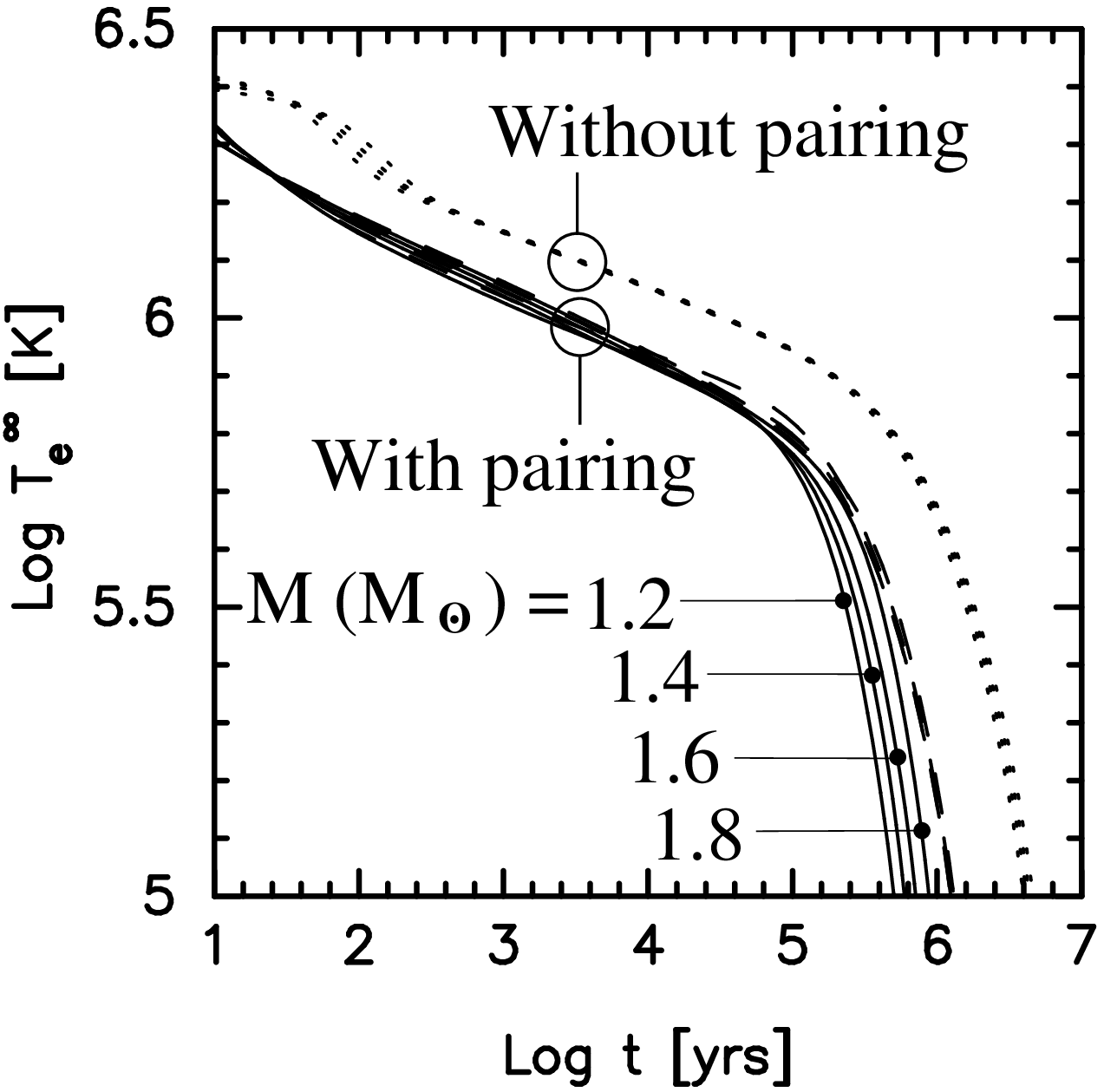}
\caption{Minimal cooling.  Left panel: comparison of cooling
         trajectories with vanishing $n$ ${^3P_2} - {^3F_2}$ gaps, labeled
         ``0'', and the three model gaps ``a'', ``b'', and ``c'' (see
         Figure~\ref{fig:Tc}).  
         Right panel: 
         dependence of the cooling trajectory on the stellar
         mass without pairing, where the mass effect is almost
         indistinguishable, and with only proton ${^1S_0}$ pairing
         (dashed curves) and both proton ${^1S_0}$ and neutron ${^3P_2} -
         {^3F_2}$ pairing.  Results from Ref.~\cite{PLPS04}.
\label{fig:cool_3P2_Mass}}
\vspace{+0.8cm}
\end{figure}

This paradigm is studied in great detail in Ref.\ \cite{PLPS04}.  Its
tenets immediately imply strong constraints on the supernuclear
equation of state, resulting in stellar radii between 11 to 12 km for
a 1.4 \Msun mass and between 9 to 10 km at the maximum mass (the
latter being not so strongly constrained, around 1.7 up to 2.3
\Msunend).  Neutrino emission from the core comprises the neutron and
proton branches of the modified Urca processes and the similar, but
less efficient, $n$-$n$, $n$-$p$, and $p$-$p$ bremsstrahlung processes (see
Table~\ref{tab:emis.core}), which are inherited from the old
``standard'' scenario.  However, essential and unavoidable within the
minimal scenario is the occurrence of nucleon pairing.  The resulting
suppression of both the nucleonic specific heat and the above mentioned
neutrino emission processes is well known, but a result is that
neutrino emission by the PBF process is dominating for most cases of
pairing gaps.  This is illustrated in Figure~\ref{fig:lum_nu}: the
left panel clearly shows that for $T_\mathrm{c}$'s of the order of $10^9$ K,
as in the cases ``a'' and ``T72'', the PBF process is more than one order 
of magnitude more efficient than the
{\em unsuppressed} modified Urca and bremsstrahlung processes when
$T$ is below $T_c$.  
Once the pairing suppression of these last two processes are taken into 
account the PBF process can
dominate by up to two orders of magnitude as shown in the right panel
of Figure~\ref{fig:lum_nu}. 
The left panel of
Figure~\ref{fig:cool_3P2_Mass} compares the effects of various neutron
${{^3P_2}} - {{^3F_2}}$ gaps, confirming the results of the previous
figure that $T_\mathrm{c} \sim 10^9$ K provides the strongest PBF neutrino
cooling, while the right panel consider the effect of the stellar
mass, which turns out to be
rather weak.  Only in case of extensive nucleon pairing can we find
some small mass dependence.

\begin{figure}
\begin{center}
\includegraphics[scale=0.65]{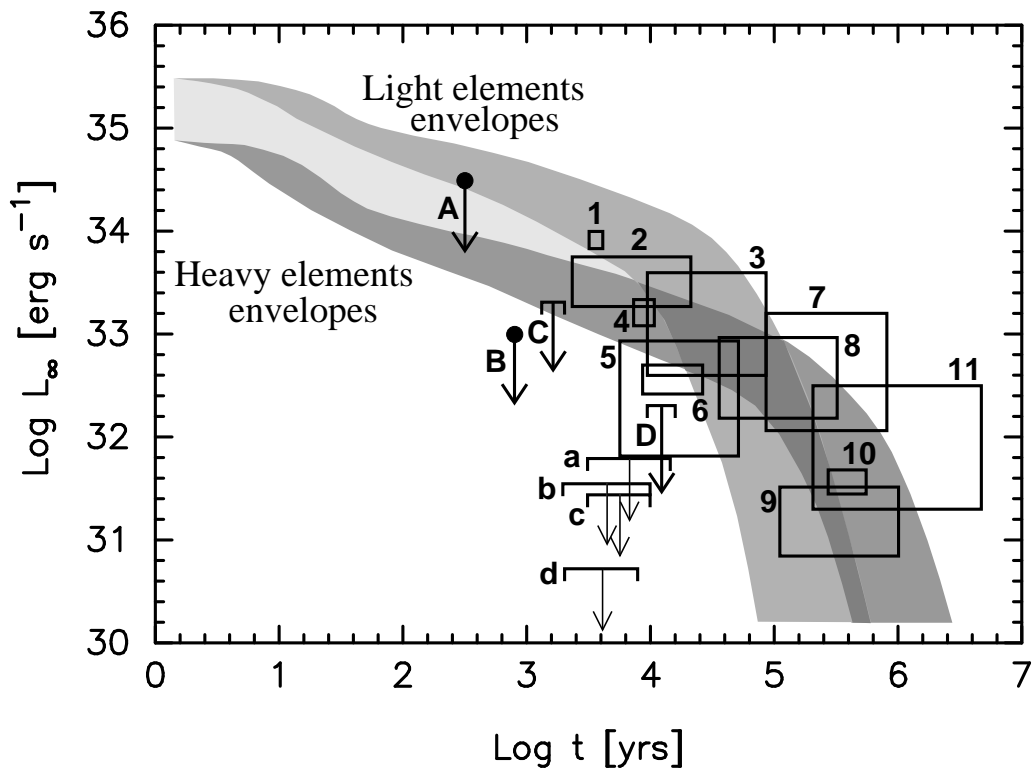}
\caption{Comparison of the predictions of the Minimal Model of neutron
         star cooling with the best presently available data.
         The two dark grey shaded areas correspond to models having
         a heavy elements envelope or an envelope with a maximum amount
         of light elements, as labeled, and the light grey area indicate
         intermediate trajectories corresponding to intermediate amounts
         of light elements in the envelope.
         The spread of predictions, for each envelope type, corresponds
         to different assumptions about the extend of nucleon (neutron  
         and proton) pairing.
         Boxes correspond to neutron stars where surface thermal emission
         is clearly detected and which have been studied in detail:
         1 to 6 are obtained from spectral fits with magnetized hydrogen
         atmospheres while 7 to 11 are from blackbody fits.
         These stars are:
         1 - RX J0822--4247 (in SNR Puppis A),
         2 - 1E 1207.4--5209 (in SNR PKS 1209--52),
         3 - PSR 0538+2817,
         4 - RX J0002+6246 (in SNR CTB 1),
         5 - PSR 1706--44,
         6 - PSR 0833--45 (in SNR ``Vela''),
         7 - PSR 1055--52,
         8 - PSR 0656+14,
         9 - PSR 0633+1748 (``Geminga''),
         10 - RX J1856.5--3754, and
         11 - RX J0720.4--3125.
         The next four stars, labeled as A, B, C, and D,
         are barely detected and in case C there is no evidence for 
         thermal emission:
         A - CXO J232327.8+584842 (in SNR Cas A),
         B - PSR J0205+6449 (in SNR 3C58),
         C - PSR J1124--5916 (in SNR G292.0+1.8), and,
         D - RX J0007.0+7302 (in SNR CTA 1).
         The last four data points, a, b, c, and d, are from 
         deep observations of four shell SNRs, considered as products of
         core collapse supernovae, in which there is no evidence of any kind
         for the presence of a compact object; they may correspond to very
         cold neutron stars or isolated black holes:
         a - ? (in SNR G315.4--2.3),
         b - ? (in SNR G093.3+6.9),
         c - ? (in SNR G084.2--0.8), and,
         d - ? (in SNR G127.1+0.5).
         See Ref.~\cite{PLPS04} for discussion and references.
}         
\label{fig:minimal}
\end{center}
\end{figure}

Comparison of the predictions of the Minimal Cooling paradigm with
the data is shown in Figure~\ref{fig:minimal}.  Predictions assuming a heavy
element or a light element envelope are plotted separately and, for
each envelope model, the width in the predictions is a result of the
uncertainty in the size of the nucleon pairing gaps. Each wide grey
strip encompasses the whole range of results when all gaps shown in
Figure~\ref{fig:Tc} are used.  The small mass dependence is also
included in these results.  Objects with the best data, shown as boxes
in the figure, show a remarkably good agreement with the theoretical
results of the minimal cooling.  One may still focus on two stars, PSR
1055--52 and RX J0720.4--3125, which may be warmer than predicted and
could be cases in which some internal heating mechanism is at work.
On the other side, PSR 0833--45 (``Vela'') and PSR 1706--44, may be too
cold and require some enhanced neutrino emission.  However, much
stronger cases for the necessity of enhanced neutrino emission are the
two pulsars PSR J0205+6449 and RX J0007.0+7302 which are clearly below
any of the predictions of the minimal cooling paradigm.  Finally, the
upper limits on the luminosity of the neutron star which may be
present in the four SNRs marked as ``a'', ``b'', ``c'', and ``d'' in
this figure doubtlessly require enhanced neutrino emission in case any
of these SNRs contained a neutron star.
Similar results has been recently obtained in Ref.~\cite{GKYG04} which,
with {\em ad-hoc} gaps, showed that even PSR J0205+6449 and RX J0007.0+7302
could be accommodated within the minimal paradigm.

\section{Enhanced Cooling}
\label{sec:enhanced}

\begin{figure}[t]
\begin{center}
\includegraphics[scale=0.40]{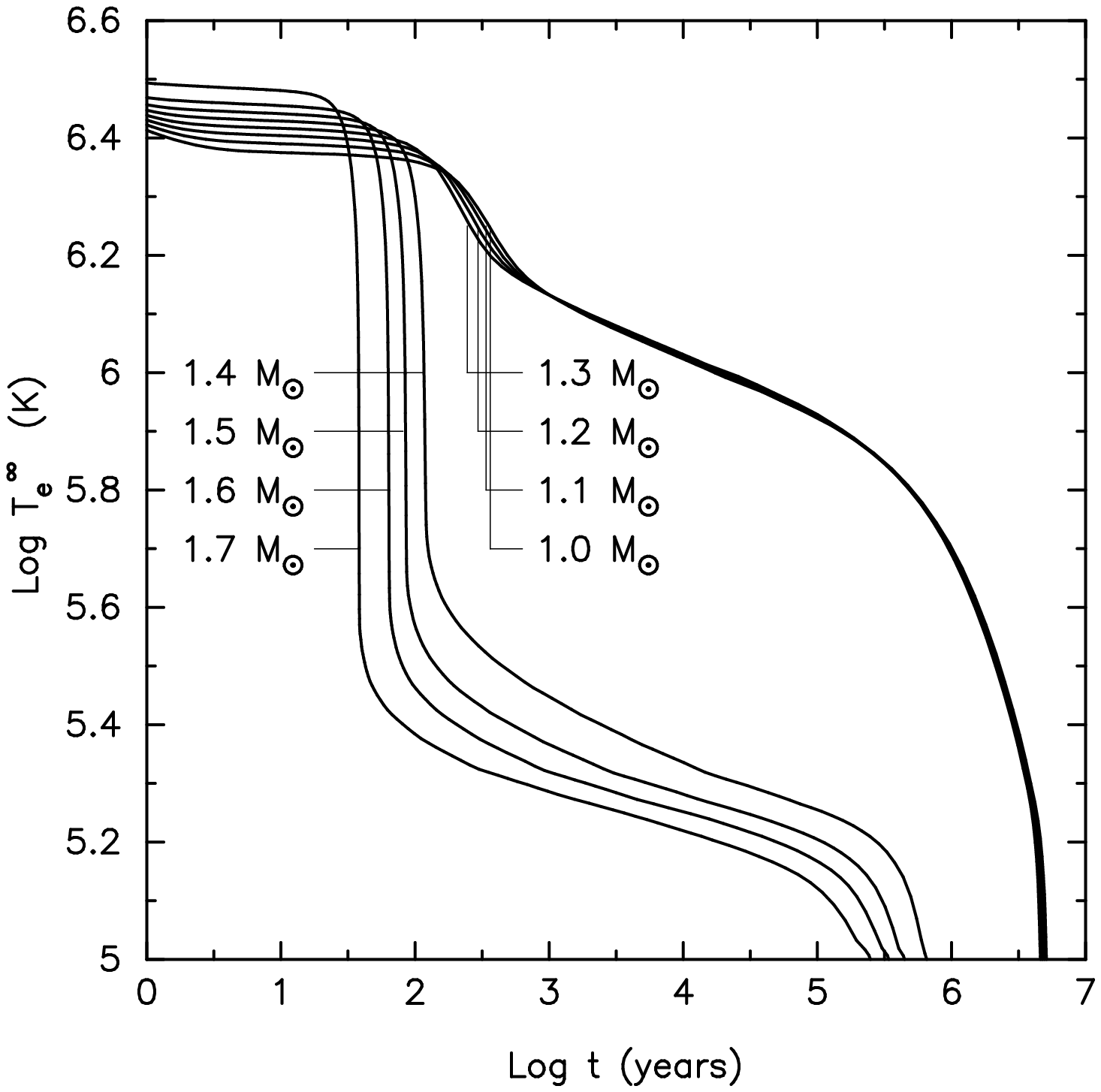}   \hfill
\includegraphics[scale=0.30]{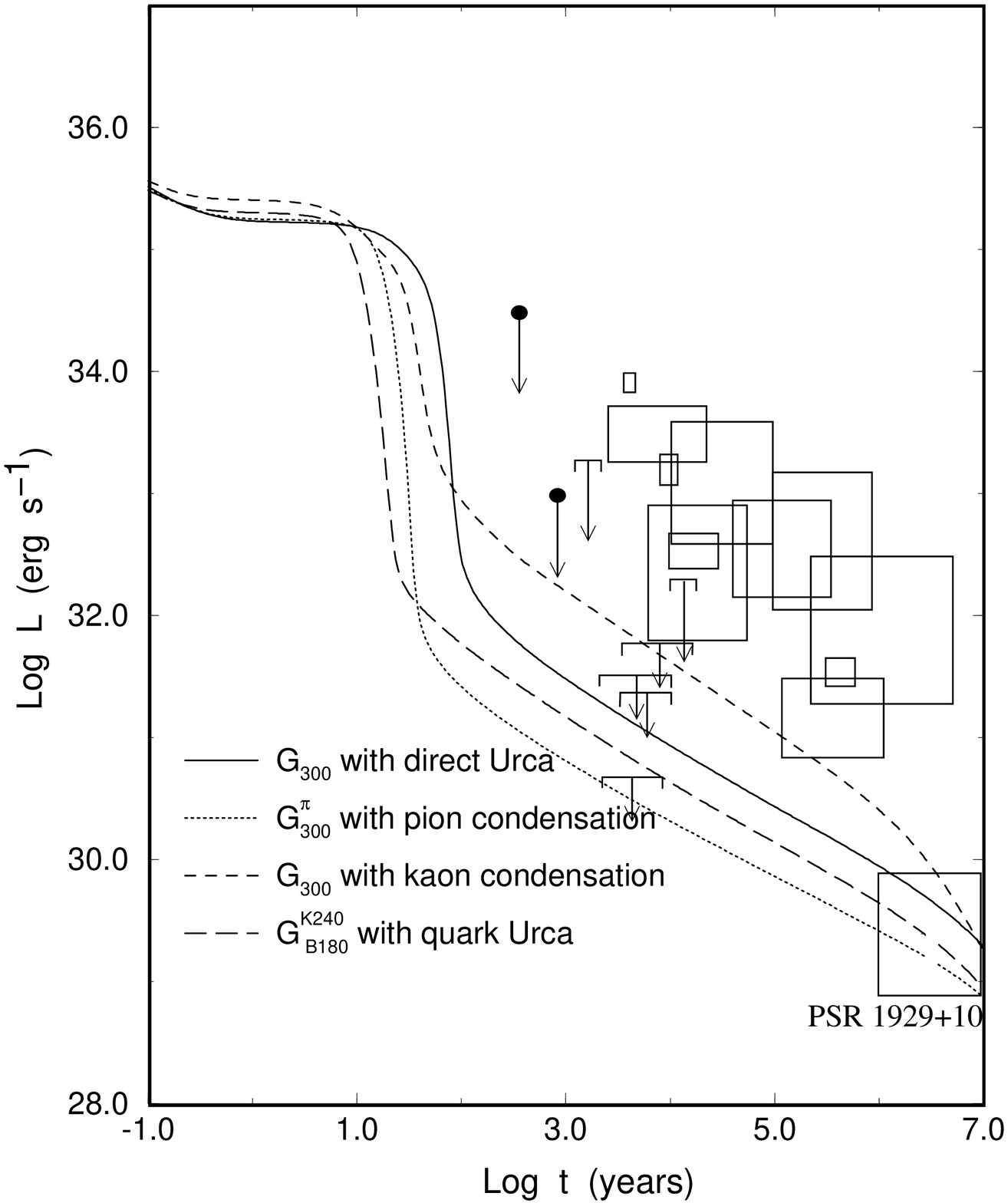}
\caption{Left panel:
         influence of the direct Urca process on the cooling of stars of various
         masses. (Figure from Ref.~\cite{PA92}.)
         Right panel: 
         influence of  different enhanced 
         neutrino emission processes on the cooling of neutron stars of mass
         $M=1.4\,M_\odot$, except for the kaon-condensed model whose mass is
         $M=1.8\,M_\odot$. (The $1.4\,M_\odot$ model is not dense enough to
         support a kaon condensation). (Figure\ from Ref.\  \cite{schaab95:a}.)
         Upper limit on PSR 1929+10 thermal luminosity is from Ref.~\cite{PSC96};
	 see Figure~\protect\ref{fig:minimal} for identification of other data.}
\label{fig:enhanced} 
\end{center}
\end{figure}

In the core, almost any chemical composition beyond the one of the minimal scenario
will open channels for enhanced neutrino emission which possibly results
in extremely fast cooling of the neutron star. 
The left panel of Figure\ \ref{fig:enhanced} illustrates the effect of the direct 
Urca process from nucleons in a model where this process is allowed for masses
above 1.35 \Msun (this specific critical mass is of course very model dependent).
Notice that the 1.4 \Msun star has an inner ``pit'' where the direct Urca
process is occurring with a mass of only 0.038 \Msun, while the 1.7 \Msun star's
``pit'' is above 1 \Msun.
After the early phase, at ages a few tens of years, the rapid temperature drop
in fast cooling stars is due to the finite thermal relaxation time of the crust
and heavier stars, having thinner crusts, relax faster.
The right panel of this figure compares the impact of the direct Urca
process on stellar cooling with other enhanced cooling processes,
which originate from pion condensation, kaon condensation and the
direct quark Urca process.
Differences between these models come as much from the various efficiencies 
for neutrino emission as from the various critical densities at which enhanced
emission becomes allowed.
Comparison with the data plotted here shows that any of these fast cooling
processes results in stars with temperature clearly incompatible with observations.

\begin{figure}[t]
\begin{center}
\includegraphics[scale=0.30]{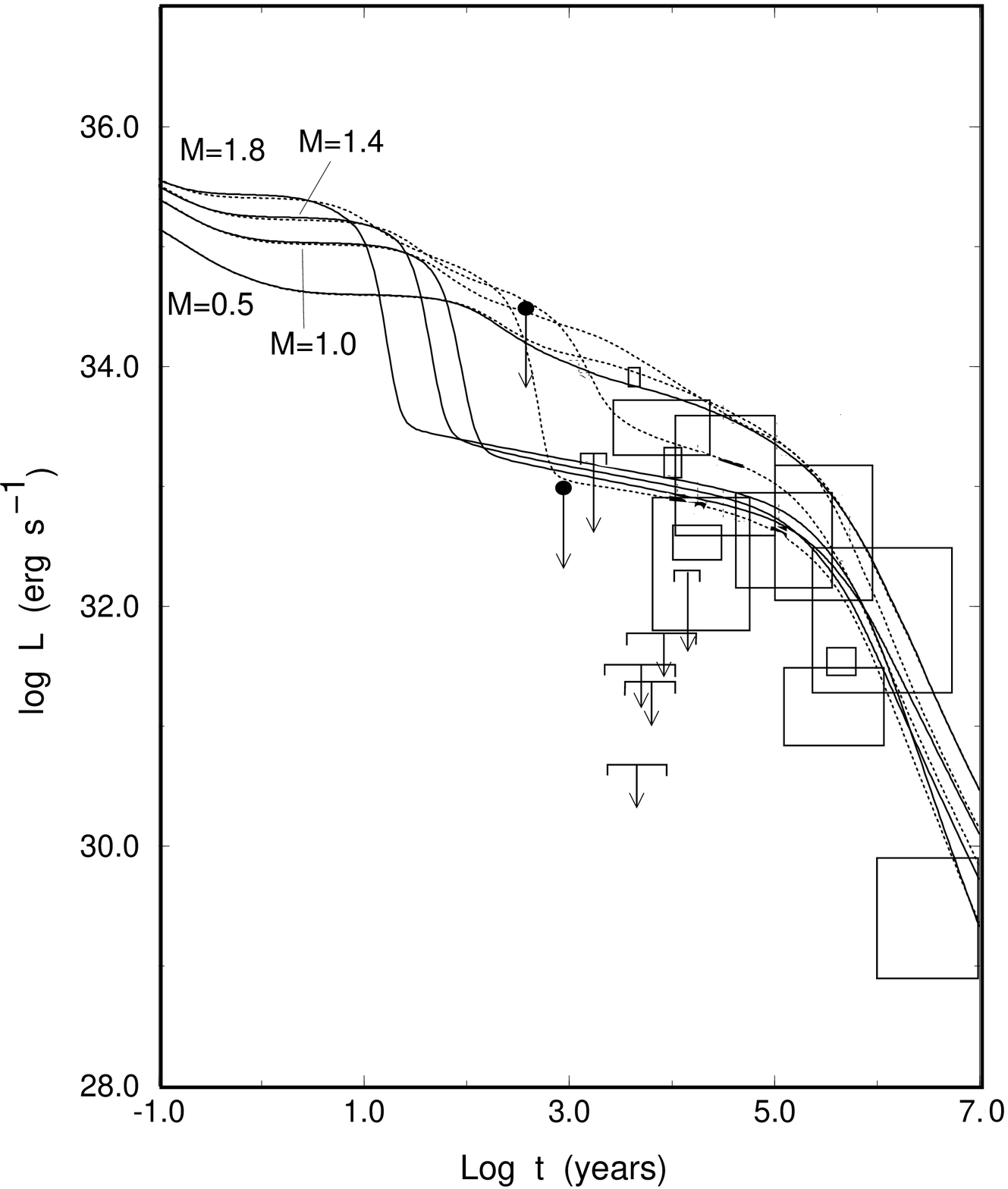}   \hfill
\includegraphics[scale=1.25]{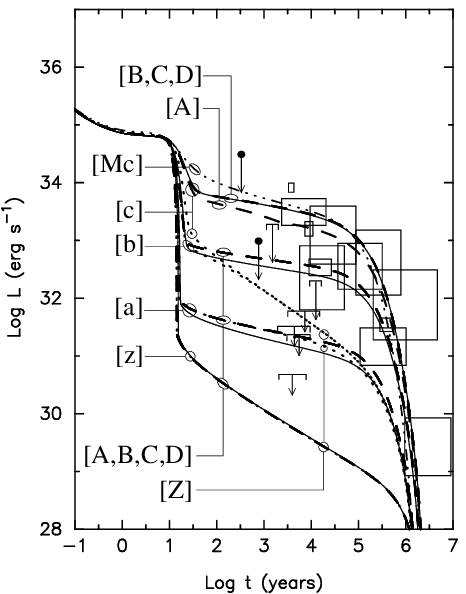}
\caption{Left panel: influence of superfluid ${^3P_2} - {^3F_2}$ neutrons
         on the cooling curves labeled $\egthpi$ and $\egth$ (with
         kaon condensation) in Figure~\ref{fig:enhanced} for various
         neutron star masses. (Figure\ from Ref.\ \cite{schaab95:a}.)
         Right panel: effect of the neutron and quark gap size.  See
         text for description and discussion.  (Figure from
         Ref.~\cite{PPLS00}.)
         The large differences in luminosities at late times between 
         the two panels are due to small differences in treatment of the
         envelope and specific heat, as discussed after Eq.~\ref{Eq:photon-cooling2}.
         See Figure~\protect\ref{fig:minimal} and \protect\ref{fig:enhanced} 
         for identification of data.}
\label{fig:enhanced+sf} 
\end{center}
\end{figure}

This naive picture is strongly affected when baryon pairing is taken
into account \cite{PB90,P89,PA92}.  As discussed in
\S~\ref{sec:pairing} the development of an energy gap induces a
suppression of the neutrino emission for any process in which the
paired component participates.  As can be seen from the left panel in
Figure\ \ref{fig:enhanced+sf}, neutron superfluidity (${^3P_2} - {^3F_2}$ in
the present example with a small superfluid gap of 0.3~MeV
\cite{SWWG96}) delays cooling significantly and moves the theoretical
temperatures right up to the region where most observed data are
concentrated.  Similar results are obtained with all enhanced cooling
scenarios as, e.g., nucleon direct Urca \cite{PA92}, $\pi^-$ condensates
\cite{UNTMT94,TTTTT02}, hyperons \cite{P98,SBSB98,PPLS00} or 
hyperons with deconfined quarks \cite{PPLS00}.  
With a high enough value of $T_\mathrm{c}$ the
enhanced neutrino emission may not even have time to act and result in
stellar temperatures as high as the ones obtained within the minimal
scenario \cite{PA92}.

The right panel of Figure~\ref{fig:enhanced+sf} illustrates several of
these considerations \cite{PPLS00}.  This panel shows cooling curves
for models with only nucleons (thin continuous curves) and with
nucleons and an hybrid phase of nuclear+deconfined quark matter
(dashed curves and dotted curves) and various hypotheses about
pairing.  Nucleon and quark direct Urca processes are allowed, except
in the case marked as ``Mc'' where they are arbitrarily turned off for
illustration.  Four different neutron ${^3P_2} - {^3F_2}$ gaps are
considered, labeled as ``z'' (zero gap), and ``a'', ``b'', and ``c''
according to Figure~\ref{fig:Tc}.  For models including quark matter in
the inner core, five quark gaps are also considered, labeled as ``Z''
(zero gap), ``A'', ``B'',''C'', and ``D'' corresponding, to maximum
gap sizes of 0.1, 1.0, 10, and 100 MeV, respectively (all flavors and
color of quarks are assumed to have the same gap for simplicity).  
Considering models without quarks matter (continuous lines),
one sees that models with a large neutron ${^3P_2} - {^3F_2}$ gap ``c'' have
almost the same evolution as the model with no direct Urca process,
``Mc'', and are all compatible with the highest measured neutron stars
temperatures.  
Models with neutron gap ``b'' could explain, as in the
left panel, the intermediate temperature neutron stars, in which case 
the warmer one should be understood as having lower masses and no direct 
Urca process allowed (model ``Mc'').  
Models with smaller neutron gaps produce too low temperatures but would 
correspond to the four upper limits (data points labeled as ``a'', ``b'',
``c'', and ``d'' in Fig~\ref{fig:minimal}).
Finally, hybrid models with quarks (dashed and dotted lines) give almost
indistinguishable results when the quark gaps are large enough 
(curves labeled ``A'', ``B'', ``C'', and ``D'')
and it is only in case of a vanishingly small quark gap that their presence
is noticeable.

As can bee seen from this brief presentation, we have an
embarrassingly large number of possible scenarios in case enhanced
neutrino cooling is occurring.  Most models have to include in an
essential way the various possible pairing gaps, about which very
little is known at high densities, to reconcile enhanced neutrino
cooling with data.  Nevertheless there are possibilities to have a
smooth transition from slow to fast neutrino cooling through strong
medium effects and without invoking pairing \cite{V01,SVSWW97}, or
with a significant amount of internal heating as discussed in
\S~\ref{sec:heating}.  Obviously, much work is still needed to
determine which scenarios are the more plausible.

\section{Cooling of Strange Quark Stars}
\label{sec:ss}

\subsection{Strange Stars with Nuclear Crusts}

\begin{figure}[tb]
\includegraphics[scale=0.30]{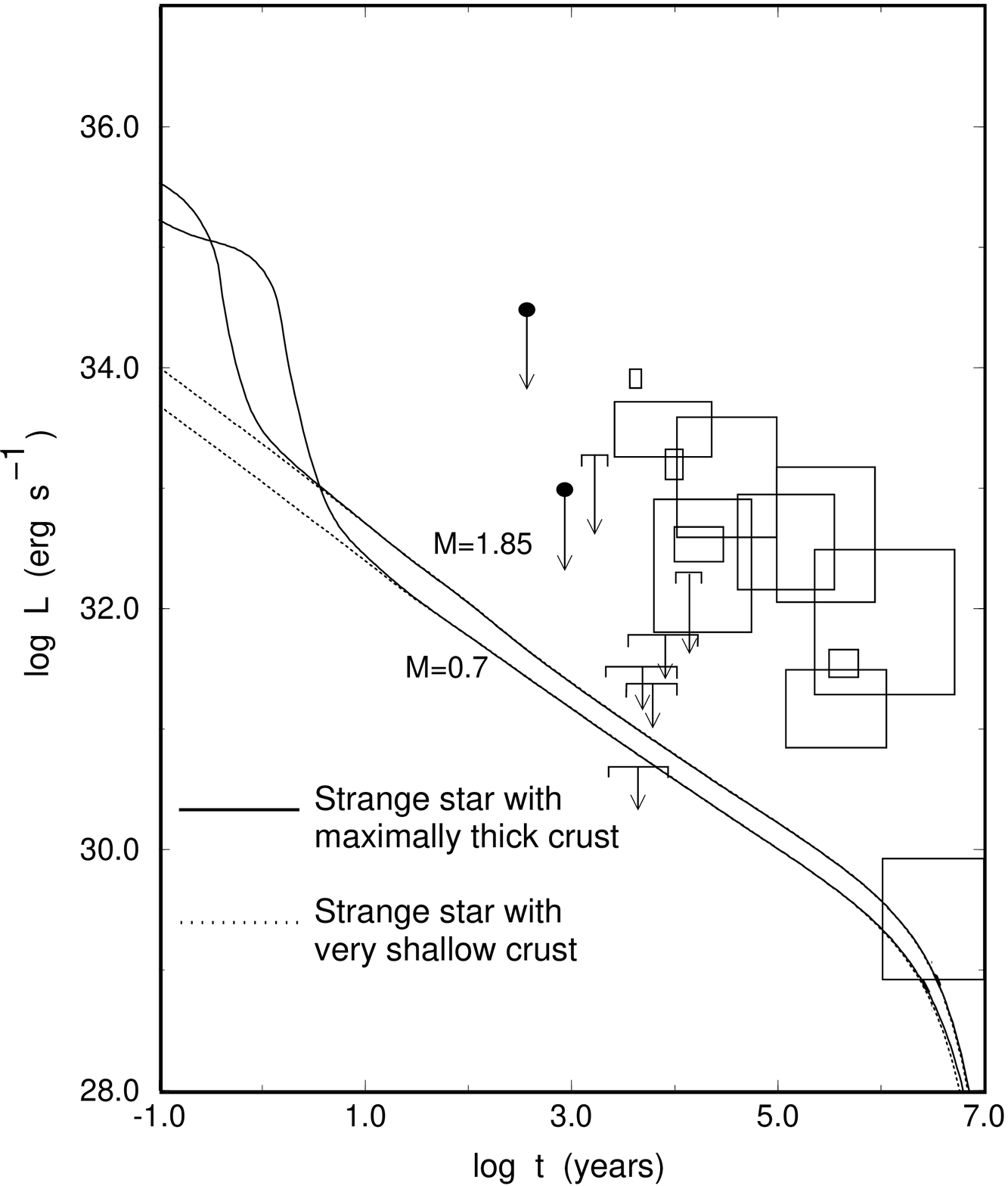} \hfill
\includegraphics[scale=0.34]{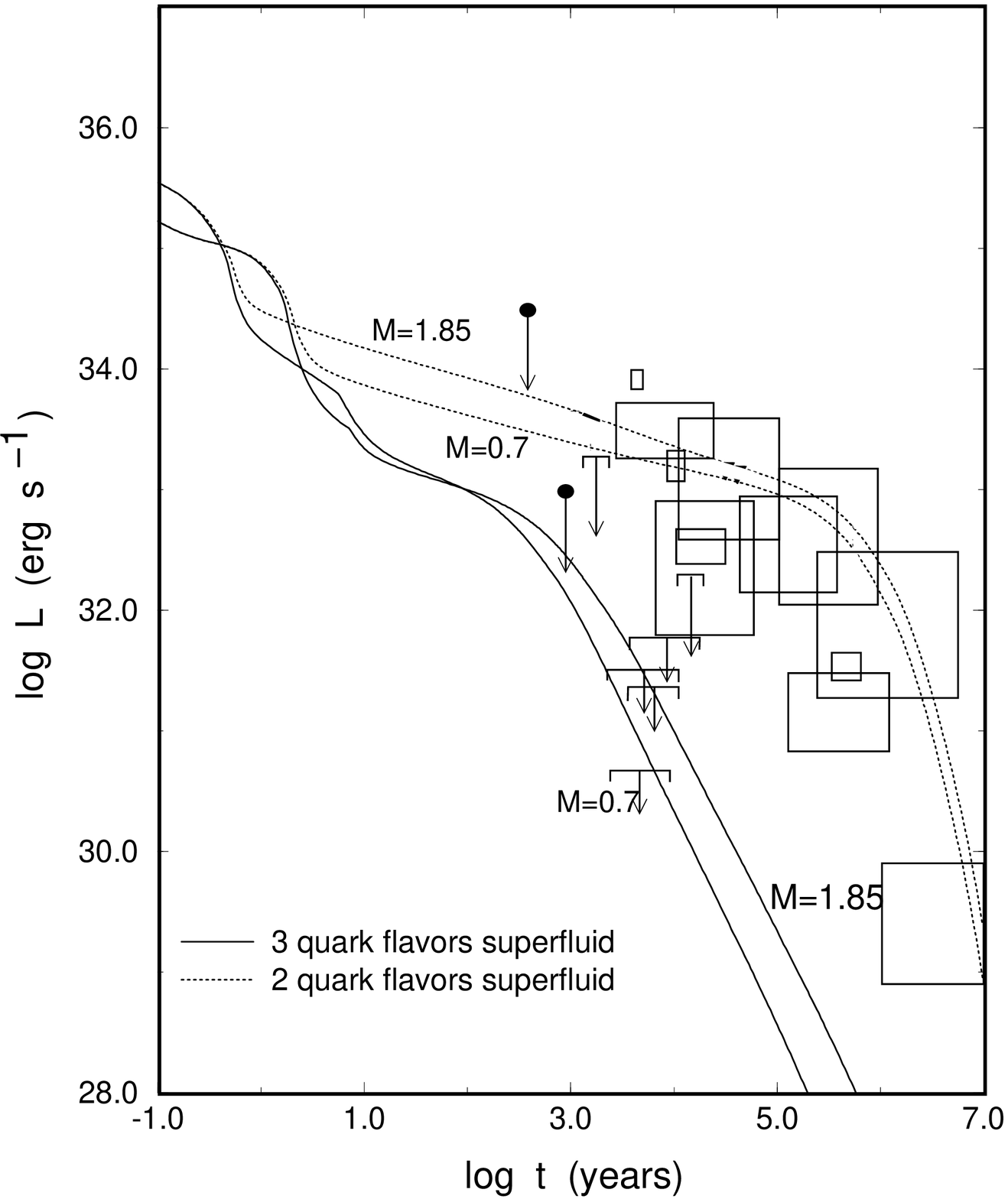}
\caption{Cooling behavior of strange stars with maximally thick crusts and
         with very thin crusts, i.e., with only an envelope
        (Figure\ from Ref.\ \cite{schaab95:a}.)
         See Figure~\protect\ref{fig:minimal} and \protect\ref{fig:enhanced} 
         for identification of data.}
\label{fig:g300_enh}
\end{figure}

If strange quark matter were in fact the true ground state of the
strong interaction (see \S~\ref{sec:EOS}), new classes of compact
stars should exist which range from dense strange stars to strange
dwarf stars \cite{weber93:b,glen94:a}. They would form distinct and
disconnected branches of compact stars and are not part of the
continuum of equilibrium configurations that include ordinary white
dwarfs and neutron stars
\cite{weber04:topr}. Figure~\ref{fig:g300_enh} shows the cooling
behavior of such strange stars with and without nuclear crusts.
One sees that not even the thickest, theoretically possible crust
(inner crust density equal to neutron drip density) does prevent
strange stars from cooling very rapidly \cite{Page92:a}. If one treats
the quarks as superfluid particles, assuming a small
density-independent gap of just 0.1~MeV, radiation of neutrinos is
greatly reduced and the stellar cooling behavior is rather similar to
that of conventional neutron stars. The different condensation
patterns of color superconducting quark matter, discussed in
\S~\ref{sec:CSC}, do not alter this picture dramatically.  In the CFL
phase, for instance, all quarks have very large gaps, $\Delta \gg T$,
so that both $\epsilon_\nu$ and $c_\mathrm{v}$ are so strongly reduced that
quark matter that may exist in the center of a neutron star would be
rendered invisible. Notice that, in spite of the CFL phase being
charge neutral by itself in bulk \cite{RW01}, a strange star in the
CFL can still support a nuclear crust when surface boundary conditions
are properly taken into account \cite{U04}.  The cooling behavior of
such stars would then be determined by the nuclear matter surrounding
the CFL quark matter core. This conclusion is strengthened by the
studies performed in Refs.\ \cite{jaikumar02:a,shovkovy02:a}.  The
cooling behavior of compact stars with 2SC quark matter in their cores
is simplified by the fact that up and down quarks may pair with a gap
$\Delta \sim 100$~MeV which is orders of magnitude larger than the
stellar temperature, $\ls 1$~MeV, and are therefore inert with respect
to the star's temperature evolution. For 2SC quark matter, however,
there exist also quark pairing channels that lead to weak pairing with
gaps on the order of several keV to about 1~MeV, which is on the same
order of magnitude as the star's temperature. These quarks may thus
not pair but, instead, radiate neutrinos rapidly via the quark direct
Urca process (Table~\ref{tab:emis.quark}). If this is the case, the 2SC
quark matter core would cool rapidly and determine the cooling history
of the star \cite{rajagopal01:a,blaschke04:a}. Naturally the
cooling behavior of such stars depends rather sensitively on the value
of the superfluid gap \cite{blaschke04:a}.

\subsection{Strange Stars with a Bare Quark Surface}

Even though it is in principle possible to cover the strange quark
matter by a layer of normal nuclear matter, as assumed in the previous
subsection, it is not clear at all how such a nuclear crust may be
formed.  When a strange star is born its internal temperature is
likely of the order of a few times $10^{11}$ K, as in a standard
proto-neutron star formed in core-collapse, and the resulting
neutrino flux is so high that it should easily be able to expel all
baryonic matter surrounding the quark matter \cite{WB92}. As a result
a new-born strange star is most certainly {\em bare}, i.e., its
surface consists directly of quark matter with no nuclear component
above it.  Such a strange star remains bare as long as its temperature
is above $\sim 3 \times 10^7$ K \cite{U97}.

If this is the case, the thermal evolution of a bare strange star would
be radically different from the one of a neutron star or a strange
star with a crust.  The surface density of the quark matter is about
$4 - 8 \times 10^{14}$~g~cm$^{-3}$ and the resulting plasma frequency
of the order of 20 MeV
\cite{alcock86:a} which may lead one to believe that a bare strange star
is unable to emit thermal radiation once it's temperature dropped
below about $10^{10}$ K, i.e., a few seconds after its birth: a bare
strange star would be a silver sphere instead of a blackbody emitter.
However this naive picture neglects the presence of electrons within
the quark matter which are bound to the quark matter by Coulomb
forces.  As such, the electrons slightly leak out of the quark matter,
by a distance which is roughly given by their Debye screening length,
producing thus an electrosphere with a thickness of a few thousand
Fermis \cite{KWWG95}.  As noted by Usov \cite{U98} the resulting
electrostatic field, of the order of $5 \times 10^{17}$ V cm$^{-1}$
\cite{alcock86:a}, is well above the Schwinger critical field of QED
and can induce copious production of ${e^-} - {e^+}$ pairs at the surface.  
However, this pair emission does not occur in vacuum, but within the
electrosphere and it is ultimately limited by electron degeneracy.
Nevertheless the resulting luminosity is enormous as long as the
surface temperature is above $\sim 10^9$ K \cite{U01}.  Pairs
outflowing from the stellar surface mostly annihilate into photons in
the vicinity of the strange star \cite{U01} resulting in a fireball of
$e^-$, $e^+$ and $\gamma$'s.  At luminosities $> 10^{41}$ erg s$^{-1}$
this plasma is optically thick and produces a blackbody spectrum while
at decreasing luminosities the spectrum progressively evolves into a
very wide annihilation line
\cite{AMU}.  At lower temperatures, when pair production become
suppressed by electron degeneracy, photons can still be emitted by
${e^-} - {e^+}$ bremsstrahlung within the electrosphere \cite{JGPP04} and
result in significant luminosities.

\begin{figure}[ht]
\includegraphics[scale=0.75]{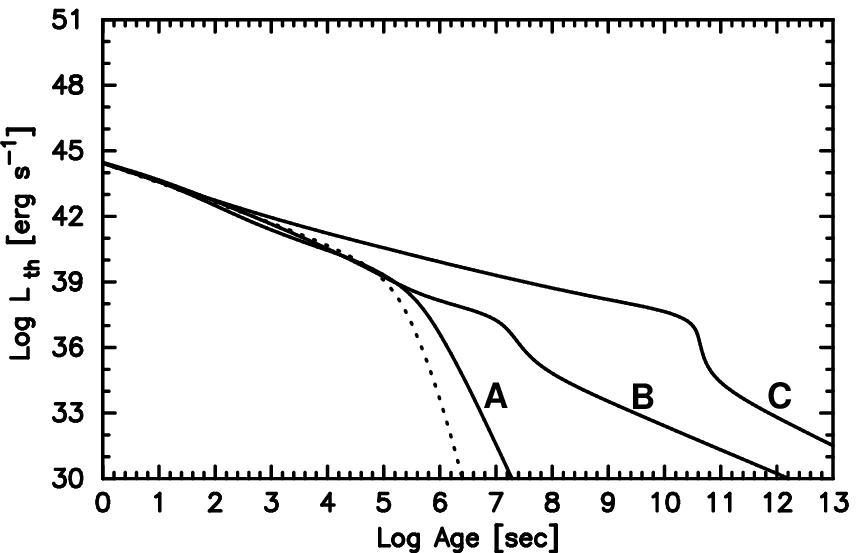} \hfill
\includegraphics[scale=0.75]{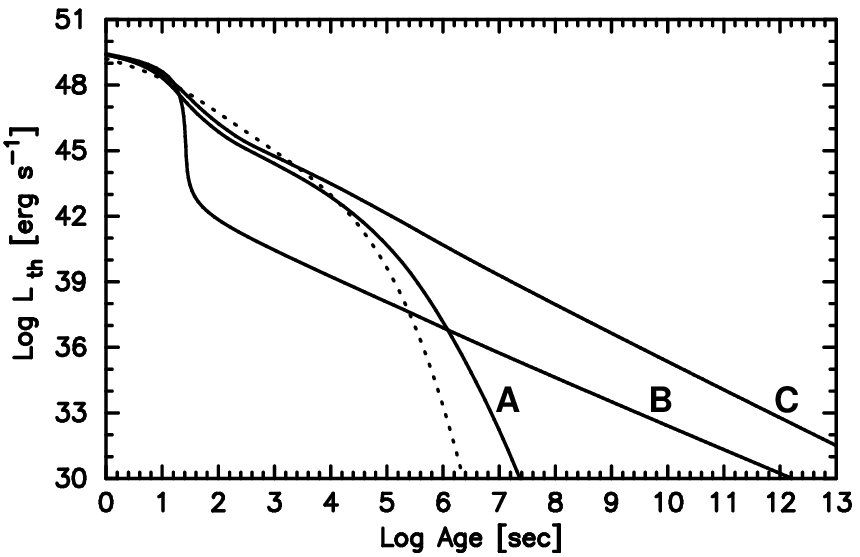}
\caption{Cooling behavior of a bare strange star,
         showing the thermal luminosity from the Usov-Schwinger
         pair-production mechanism vs. age.  In the left panel it is
         assumed that heat, within the star, flows to the surface by
         diffusion while in the right panel convection is allowed
         resulting in much stronger heat flow and higher thermal luminosities.  
         In the dotted curves
         no quark superconductivity is included while the continuous
         curves labeled A, B, \& C have quark color superconductivity
         taken into account for three possible scenarios (see text for
         details). (From Ref.\ \cite{PU02}.)  }
\label{fig:SS_bare}
\end{figure}

The thermal evolution of a bare strange star with the Usov-Schwinger
pair production mechanism has been studied numerically in \cite{PU02}
and results are shown in Figure~\ref{fig:SS_bare}.  Uncertainties about
the pairing state of quark matter are taken into account schematically
in the three scenarios marked in the figure as ``A'', ``B'', and
``C'', in which the gaps suppress neutrino emission with increasing
efficiency, leading thus to slower cooling of the star.  These results
show that bare strange stars can produce extremely high thermal,
luminosities, well above the Eddington limit, over extended periods of 
time, but with considerable 
uncertainty due to the precise phase of color superconductivity
present at these ``low'' astrophysical densities.  Notice, however,
that the photon emission from ${e^-} - {e^+}$ bremsstrahlung in the
electrosphere \cite{JGPP04}  was not included in these models, and will
significantly change the evolution when the thermal luminosity has dropped
below $10^{40}$ erg s$^{-1}$.

\section{Magnetic field effects in the crust} 
        \label{sec:magcrust}

All the heat stored in the core of the neutron star and eventually
irradiated away from its surface by photons has to be transported
through the crust.  In the absence of rotation and magnetic field this
transport in the crust is spherically symmetric.  
While the effects of rotation are quite small even for
millisecond pulsars, the presence of magnetic fields may cause
significant deviations from the spherical symmetry of the transport
processes, even for quite ``standard'' field strength of $\sim 10^{12}$G.  
Due to the classical Larmor rotation of electrons, a magnetic field 
causes anisotropy of the heat flux since the heat conductivity becomes 
a tensor whose components perpendicular, $\kappa_\perp$, and parallel,
$\kappa_\|$, to the field lines become
\be
\kappa_\perp = \frac{\kappa_0}{1+(\Omega_B \tau)^2}
\;\;\;\;\;
\mathrm{and}
\;\;\;\;\;
\kappa_\| = \kappa_0
\ee
where $\kappa_0$ is the conductivity in absence of magnetic field,
$\Omega_B$ the cyclotron frequency of the electrons and $\tau$ their 
collisional relaxation time.
In the neutron star crust $\Omega_B \tau \gg 1$ is easily realized
and, as a result, heat flows preferentially along the magnetic field 
lines \cite{GKP04}.
This effect is moreover amplified at the
surface by the well known non-isotropy in the envelope as described in
\S~\ref{sec:envelope}.  

\begin{figure}[t]
\includegraphics[width=13.0cm]{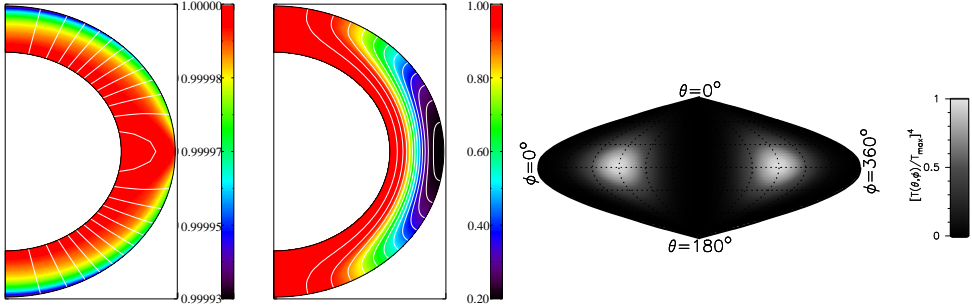}
\caption{Temperature distribution and magnetic field lines in the crust 
for a ``core'' (left panel) and a ``crustal'' (central panel) dipolar field 
(from Ref.\cite{GKP04}), and surface temperature distribution resulting from
the ``crustal'' field (right panel).
The thickness of the crust in left and central panels has been stretched by a factor 5
for clarity and the two bars show the temperature scales in units of 
$T_\mathrm{core}=10^6$ K. The ``core'' field gives a surface temperature
distribution very similar to the left panel of Figure~\protect\ref{fig:Tdistr1}.
For the ``core'' filed on has $T_e = 1.15\times 10^5$K while for the
``crustal'' field $T_e = 0.89\times 10^5$K.
In both cases, the field strength is $B_0=3\times 10^{12}$~G.
(The neutron star considered here has a 1.4 \Msun mass and a radius of 11 km.)}
\label{fig:heat_field}
\end{figure}

The non-isothermality of the subjacent crust
depends strongly on the internal geometry of the field \cite{GKP04}.  
Assuming a dipolar field structure, outside the star the radial dependence
of the field is uniquely determined, $\propto r^{-3}$, while inside the 
star it depends on the location of the electric currents, through Amp\`ere's law.
Assuming currents are exclusively located in the core, ``core field'',
the $r^{-3}$ dependence also applies to the crust while if they are
exclusively located in the crust and the field does not penetrates the core,
``crustal field'', the field topology in the crust is radically distinct:
for the {\em same} external field in the latter case field lines in the crust 
are squeezed into the crustal shell and have hence a very large meridional
component.
This meridional component of the crustal field inhibits the radial
flow of heat which is hence redirected preferentially toward the magnetic poles.
This difference is illustrated in Figure~\ref{fig:heat_field}.

The drastic difference in the crustal
temperature distribution for the different field structures which are
characterized by the same dipolar field structure and strength outside
the neutron star, causes significant differences in the
surface temperature distribution which will have several observational
consequences:

\noindent 1. A non-uniform surface temperature induces a modulation of the
observed soft X-ray thermal emission \cite{P95}.  The stronger
channeling of the heat flow toward the polar regions in the case of a
crustal field will result in larger amplitude in the pulse profile.

\noindent 2. The differences in the photon luminosities for a core or a 
crustal field will also affect the long term cooling of neutron stars.
A neutron star having a magnetic field confined to its crust will stay
warmer for a longer time, due to its lower photon luminosity, compared
to a neutron star with a field penetrating its core.

\noindent 2. This may open a way to study the {\em internal} geometry of
the magnetic field. We may be able to distinguish between field geometries
where the currents are essentially localized in the crust or in the core,
or even detect the imprint of the presence of a toroidal field
\cite{PKG05}

\section{Heating Mechanisms} 
         \label{sec:heating}

The expression ``heating mechanism'' refers to the term ``$H$'' in
Eq.~\ref{equ:energy-conservation}
and generically encompasses all possible dissipative processes which will
inject heat into the star by tapping into various forms of energy:
magnetic (\S~\ref{sec:joule}), rotational or chemical (\S\ref{sec:friction}).

\subsection{Magnetic Field Decay and Joule Heating}
         \label{sec:joule}

Given that most neutron stars have strong magnetic fields,
magnetic energy is a natural reservoir from which to extract heat by the 
Joule effect from the decaying electric currents.
Assuming a uniform internal field  of strength $B=10^{13} B_{13}$ G, 
one can roughly estimate an amount
\be
E_\mathrm{mag} \sim \frac{B^2}{8\pi} \times \frac{4}{3} \pi R^3
\sim 2 \times 10^{43} \; B_{13}^2 \;\; \mathrm{erg}
\label{eq:magenergy}
\ee
of stored magnetic energy.
With a field decay time scale $\tau = 10^6 \tau_6$ yrs this gives us an
equivalent ``magnetic heating luminosity''
\be
L_\mathrm{mag} \simeq \frac{E_\mathrm{mag}}{\tau} \sim 
6 \times 10^{29} \; \frac{B_{13}^2}{\tau_6} \;\; \mathrm{erg} \; \mathrm{s}^{-1}
\label{eq:heatlum}
\ee
This simple estimate indicates that decay of a standard magnetic field
can alter the neutron star thermal evolution at late ($> 10^6$ yr) times.
In the case of a magnetar, ultra-strong fields and potential channels for
fast decay (i.e., $\tau \ll 10^6$ yrs) can nevertheless lead to significant
heating in young stars \cite{TD96,CGP00,ACT04}.

\begin{figure}[ht]
\begin{center}
\includegraphics[scale=0.60]{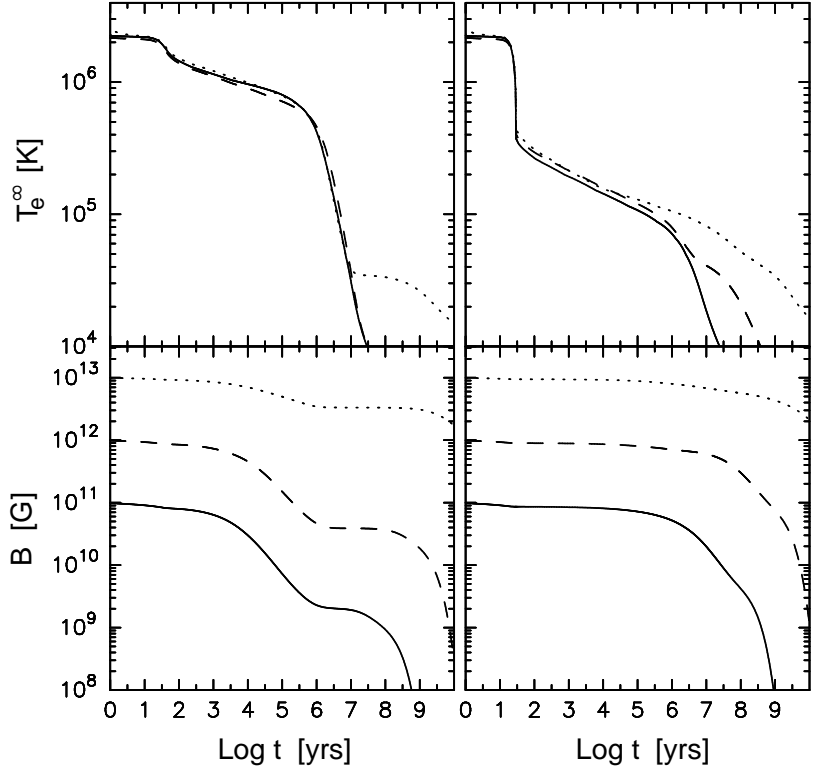}
\caption{Thermal (upper panels) and magnetic (lower panels) evolutions
of neutron star undergoing slow (left panels) and fast (right panels)
neutrino cooling, with Joule heating from magnetic field confined to
the stellar crust.
The three representative case of initial field strengths cover a wide range of 
model parameters:
$B_0=10^{11}$, $10^{12}$, and $10^{13}$ G have their supporting currents
initially concentrated at densities of
$\rho_0 \sim 10^{12}$, $10^{13}$, and $10^{14}$ gm cm$^{-3}$, respectively.
 (Adapted from Ref.~\cite{PGZ00}.)  }
\label{fig:joule}
\end{center}
\end{figure}

We will now describe some numerical results for the simplest case of a
standard magnetic field, following the Ref.~\cite{PGZ00} which
considered the case where magnetic energy is converted into heat
through Ohmic decay of the currents supporting the field.  This Joule
heating contributes to the source term $H$ in
Eq.~(\ref{equ:energy-conservation}), where the heat production by
field decay per unit of (proper) time and volume is given by ${{\vec
j}^2}/{\sigma}$, $\vec j$ being the electric current density and
$\sigma$ the scalar electric conductivity.  
Joule heating is specially effective  in the crust, which
never becomes superconductive.  Therefore, Ref.~\cite{PGZ00}
considered only magnetic fields which are, along with the currents
supporting them, confined to the neutron star crust and some results
are shown in Figure~\ref{fig:joule} (see also Ref.~\cite{UK97}.)
The amount of heat locally released by Joule heating is determined by 
the strength of the field and its decay rate at that time.  
Most of the magnetic energy is dissipated at early times when it has 
almost no effect because the star's thermal energy is still too large.
It is only when the star has become sufficiently cold that $L_\mathrm{mag}$ 
becomes significant, unless the field strength is of magnetar size.
As illustrated in Figure~\ref{fig:joule}, scenarios with a weak field supported
by currents located at low density, i.e., decaying fast while the star is still
hot, lead to virtually no observable effect.
On the contrary, scenarios with strong field supported by currents at high
densities keep a large amount of magnetic energy to be dissipated on long 
time scales, the best cases being fast cooling stars which result in the
highest temperatures at ages $\sim 10^9- 10^{10}$ years.

The above description is based on a linear evolution of the field but
non-linear effects caused by the Hall-drift in the highly 
magnetized crust may lead to an instability, which transfers magnetic 
energy rapidly from large scale structures into much smaller ones
\cite{GR92,RG02}.
In that way, close to the NS surface small  scale ($\sim 0.1 - 1$ km) 
field structures, as necessary for the onset of the pulsar mechanism, 
will be created which simultaneously are sources of accelerated 
($t_\mathrm{Hall} \sim 10^4$yrs) Joule heating \cite{GRG03}.
Detailed numerical modeling of the effect of this instability on the
cooling of the pulsar remains to be performed.

\subsection{Dissipative motion of vortex lattices and readjustment to 
equilibrium}
        \label{sec:friction}

The dissipative motion of vortex lattices and the rotational readjustment 
of the stellar equilibrium structure are other key processes that
contribute to the heating of neutron stars. The relative effectiveness
of these heating processes, however, varies significantly from one
process to another, and depends sensitively on
\begin{figure}[tb] 
\begin{center}
\includegraphics[scale=0.5,angle=-90]{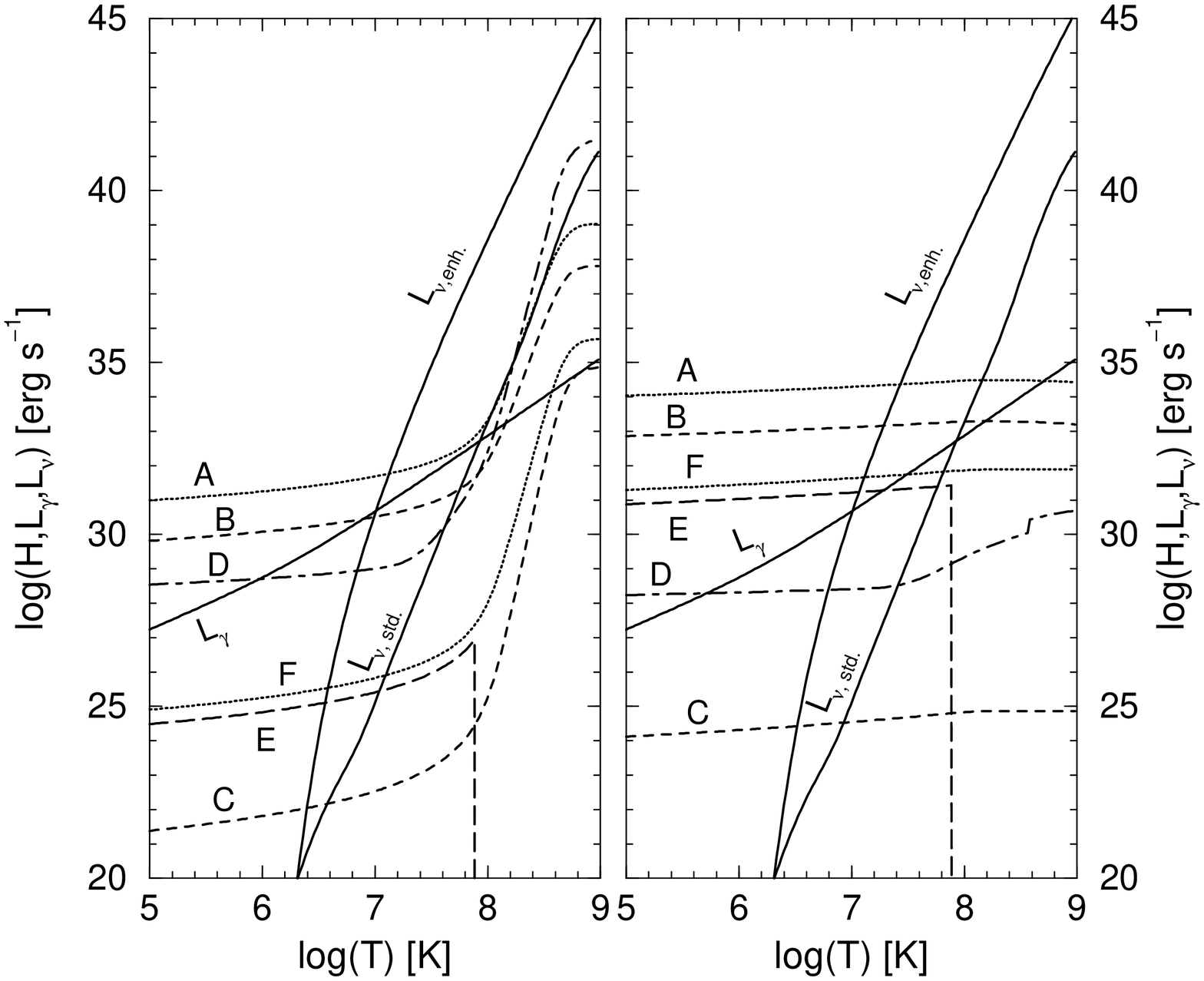}
\caption{Heating rate $H$ as a function of internal star temperature for 
the heating scenarios (A)--(F) listed in the text. The left panel
corresponds to $K \sim 10^{-15}$~s, the right panel to $K \sim
10^{-22}$~s. The photon and neutrino luminosities of standard cooling
and enhanced cooling models ($L_\gamma$, $L_{\nu,\; \mathrm{std.}}$,
$L_{{\nu},\; \mathrm{enh.}}$) are shown for comparison. (Figure\ from Ref.\
\cite{schaab99:a}.)}
\label{fig:H}
\end{center}
\end{figure} 
the value of the rotational parameter $K$, which enters the power-law
relation $\dot\Omega(t) = - K(t) \Omega^n(t)$ characterizing the
spin-down of pulsars.  Figure \ref{fig:H} shows the heating rate $H$
(which enters in Eq.\ (\ref{equ:energy-conservation})) computed for
several competing, internal heating processes and parameter
sets. These are \cite{schaab99:a}: A) vortex pinning at nuclei for the
Epstein-Baym parameter set (EB-pinning); B) vortex pinning at nuclei
for the parameter set of Pizzochero et al.; C) co-rotating crust; D)
co-rotating core; E) crust cracking; and F) chemical heating.
Chemical heating occurs because of the changing chemical composition
with stellar frequency. The matter would maintain chemical equilibrium
if the relaxation timescales for the weak reaction processes were small
compared to the timescale of rotational evolution. These timescales,
however, were found to be comparable and the stellar composition
therefore departs from chemical equilibrium, which modifies the
reaction rates \cite{haensel92:a} and leads generally to a net
conversion of chemical energy into thermal energy
\cite{reisenegger92:a,iida97:a}.  In general, internal heating leads
to enhanced stellar surface temperatures, which can be very pronounced
for rotating neutron stars of middle ($K \sim 10^{-15}$~s) and old ($K
\sim 10^{-22}$~s) ages \cite{schaab99:a} or millisecond pulsars 
\cite{R97}. As shown in \cite{schaab99:a,P97} (see also \cite{USNT93})
, the increase in surface temperature caused by
the internal heating due to thermal creep of pinned vortices and the
outward motion of proton vortices in neutron stars cooling rapidly via
one of the enhanced mechanisms discussed in \S~\ref{sec:enhanced}
would lead to a good agreement with the observed data.

\section{Cooling Neutron Stars in Soft X-Ray Transients}
        \label{sec:SXRT}

Neutron star undergoing accretion from a companion in a close binary
system can also give us information about the state of dense matter at
supranuclear densities.  In case of continuous accretion, the bulk of
the observable X-ray emission is dominated by the release of
gravitational energy when matter hits the neutron star surface and by
frictional heating in the inner part of the accretion disc, in case
where such a disc exists.  Nuclear energy is released during X-ray
bursts, and superbursts, which may teach us about the thermal response
of the stellar core \cite{RS04}.

A very interesting class of accreting neutron stars are the one in the
so-called Soft X-Ray Transients (``SXRTs'') which undergo recurrent
surges of activity separated by long phases of relative ``quiescence''
during which accretion very likely does not occur.  The substantial
X-ray brightening observed in outburst ($L_\mathrm{o} \sim 10^{37-38}$
erg s$^{-1}$) together with type I bursts unambiguously indicate that
episodes of intense accretion occur onto a stellar surface, i.e., a
neutron star, and not into a black hole.  In quiescence the faint
X-ray emission has a luminosity in the order of $L_\mathrm{q} \sim
10^{31-33}$ erg s$^{-1}$, i.e., comparable to a standard isolated
cooling neutron star.  Typically, the duration of an outburst, $t_\mathrm{o}$, 
is considerably shorter than the recurrence time between two outbursts,
$t_\mathrm{r}$.  In principle, the effects of accretion are the same as
in case it is continuous but during the quiescence phase we are most
certainly detecting thermal emission from the cooling of the neutron
star heated up by the accretion.  Given the high internal temperature
of the star, the heat released in the upper layers from the accretion
and the bursts flows back to the surface and is radiated away because
of the large temperature gradient in the envelope.  Nevertheless, the
star stays hot due to energy release from non-equilibrium process in
the interior: this heat can be stored in the stellar interior and
slowly released when accretion stops.

Non equilibrium processes certainly occur in the crust of an accreting
neutron star: thermonuclear processes at the surface will produce
iron-peak nuclei which are then pushed to higher densities.  In their
way, these nuclei will undergo electron capture, neutron emission and
absorption and eventually pycnonuclear fusions \cite{HZ99-03}.
Overall, about 1.5 MeV is released as heat for each accreted baryon.
This energy is enough to explain the observed quiescent (thermal)
luminosities $L_\mathrm{q}$ in most cases \cite{BBR98}:
\be
L_\mathrm{q} \simeq f \times Q_\mathrm{nuc} \frac{<\!\dot{M}\!>}{m_\mathrm{u}} \simeq f 
\times 6
\times 10^{32} \frac{Q_\mathrm{nuc}}{1.5 \mathrm{MeV}} \,
\frac{<\!\dot{M}\!>}{10^{-11} \Msun \mathrm{yr}^{-1}} \; \mathrm{erg} \, \mathrm{s}^{-1} ,
\label{Eq:f}
\ee
where $<\!\dot{M}\!>$ is the time average of the accretion rate,
$Q_\mathrm{nuc}$ the energy released deep in the crust per accreted
baryon, and $f$ the fraction of $Q_\mathrm{nuc}$ which is stored in the
stellar interior, i.e., not lost by neutrino emission.  
The luminosity $L_\mathrm{o}$ during an accretion outburst can be estimated as
$L_\mathrm{o} \simeq (\Delta M / t_\mathrm{o}) (GM/R)$ where 
$\Delta M$ is the mass accreted during the
outburst. Writing $<\!\dot{M}\!> \simeq \Delta M/t_\mathrm{r}$ one then
obtains, following \cite{BBR98}, 
\be
\frac{L_\mathrm{o} t_\mathrm{o}}{L_\mathrm{q} t_\mathrm{r}} 
\simeq \frac{GM}{R} \frac{m_\mathrm{u}}{f Q_\mathrm{nuc}}
\simeq \frac{200}{f} \, .
\ee
Notice that $L_\mathrm{o}/L_\mathrm{q}$ is independent of the source's
distance, which is often poorly constrained, while $t_\mathrm{o}$ and
$t_\mathrm{r}$ can in principle be directly obtained by monitoring the
source for a long enough time.

\begin{figure}[htb] 
\begin{center}
\includegraphics[scale=0.4]{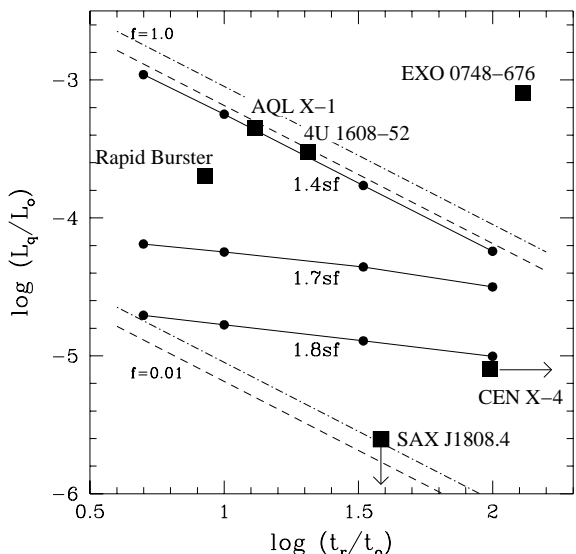}
\vspace{-0.5cm}
\caption{Quiescent to outburst luminosity ratio plotted versus the 
recurrence time over the outburst time. The filled squares represent
observed values.  Estimates for the efficiency of heat storage $f=1$
and $0.01$, Eq.~(\ref{Eq:f}) (but with minor GR corrections), are
shown for a $1.4\; \Msun$ star (dashed-dot) and $1.8 \; \Msun$ star
(dashed).  Dots connected by lines show numerical results (see text )
for different neutron star masses as labeled: the 1.4 \Msun case has
no enhanced neutrino emission, i.e., $f \simeq 1$, while the other two
cases at 1.7 and 1.8 \Msun have increasingly more efficient neutrino
emission, i.e., $f \ll 1$ (the precise value of the neutron star
masses are, however, very model dependent).  Adapted from
\cite{CGPP01}, with new data for SAX J1808.4-3658 from \cite{Cetal02}.}
\label{Fig:Lo_Lq}
\end{center}
\end{figure}

Detailed numerical calculations \cite{CGPP01,UR01} confirmed this simple
and elegant picture and are compared with data in Figure~\ref{Fig:Lo_Lq}
(see also \cite{YLH03,YLPGC04}).  Results in this figure are based on
a dense matter model in which the direct Urca process is allowed at
high densities and controlled by neutron pairing. The two stars Aql
X-1 and 4U 1608--52 show very good agreement with the simple model and
the numerical results with $f \simeq 1$, i.e., with negligible
neutrino losses and thus very probably are low mass neutron stars.
The star in Cen X-4 is noticeably off and seems to require very strong
neutrino losses.  However, it must be emphasized that in 35 years
since its discovery it has undergone only two bursts (in 1969 and
1979) and hence its recurrence time $t_\mathrm{r}$ is highly
uncertain: it is not impossible that this point should be plotted much
further to the right, as indicated by the arrow in the figure.  The
most interesting and intriguing object is certainly SAX J1808.4-3658,
dubbed as ``The accreting pulsar'', which provided the first strong
observational evidence for the evolutionary scenario that millisecond
pulsars owe their short rotational period to a long phase of accretion
in a low mass X-ray binary system.  The very low upper limit on its
thermal quiescent luminosity $L_\mathrm{q}$ \cite{Cetal02} requires, 
within the present scenario, very strong neutrino emission (even beyond
what has been considered in \cite{CGPP01} and shown in Figure~\ref{Fig:Lo_Lq},
i.e., an inner core where enhanced neutrino emission is not suppressed at all 
by neutron pairing) and hence a large mass which is in perfect agreement with
the idea that recycling of a neutron star to a millisecond rotation
period by accretion requires a mass transfer of about $0.2$ to $0.5$
\Msun \cite{CST94}.
Notice that the first precise measurement of the mass of a millisecond
pulsar, PSR J1909-3744, was recently announced, giving a value of 
$1.438 \pm 0.024$ \Msun \cite{JHBOK05}, 
at the upper edge of the mass range of binary pulsars \cite{TC99}.
(The latter, which are {\em not} millisecond pulsars, come from short-lived 
double massive star systems and did not have time to accreted a significant 
amount of mass.)
The measured mass of PSR J1909-3744 demonstrates that spinning-up of a neutron
star to millisecond period does not requires accretion of much more that
0.2 \Msun.
Other mass estimates of binary millisecond pulsars \cite{nice04:a}, though not
as accurate, do indicate that much larger masses are reachable, e.g., 
PSR J0751+1807, which has a mass lower limit of 1.6 \Msun.

Another, and still poorly explored, aspect of these objects is their
short time-scale thermal response to the accretion phases.  This is an
aspect of the problem from which extremely important information about
both the structure of the neutron star crust and the thermal state of
their core can be obtained \cite{UR01,BBC02}, and about which
intriguing observational results are being found (see, e.g.,
\cite{WHML04}). 

\vspace{-0.4cm}
\section{Epilogue and Conclusions}
        \label{sec:epilogue}
\vspace{-0.4cm}

Confrontation of the predictions of the Minimal Model with data in
\S~\ref{sec:minimal} showed a reasonably good agreement with presently
available data on isolated cooling neutron star, i.e., no compelling
evidence for the occurrence of enhanced neutrino emission is found.
Nevertheless, most models of dense matter predict the presence of some
form of exotic matter at high enough density, or at least a high
proton fraction, which allow enhanced neutrino emission.
Are these ``non-minimal'' scenarios wrong or 
are we missing cooling neutron stars ?

Before hasting into conclusions, one must first emphasize that the fact
the results of the Minimal Cooling scenario are compatible with the
data does not mean that the other scenarios are ruled out
(unless one is a fervent adept of the ``Occam's Razor Principle'').
A scientific paradigm is refuted when it is incompatible with
experimental facts, not because its competitors are compatible.
Alternative scenarios, ``exotic'' ones or simply with a high proton 
fractions, whose chemical composition allows enhanced neutrino emission 
do not necessarily imply fast cooling since the neutrino emission can 
be suppressed by pairing we showed in \S~\ref{sec:enhanced}.
With large enough gaps these scenarios can be compatible with the
data as illustrated in Figure~\ref{fig:enhanced+sf}.
Moreover, enhanced neutrino emission is expected to be allowed, if at all,
only at high enough densities, and hence for sufficiently massive neutron
stars, the critical neutron star mass depending on the particular model.

We have no direct information at all about the masses of the observed
nearby cooling neutron stars. Many measured masses of radio pulsars in
binary systems are between 1.25 to 1.44 \Msun \cite{TC99,Letal04}.
Exceptions to this are the mass of the binary neutron star J0751+1807
which is in the range $2.1^{+0.4}_{-0.5}$ \Msun \cite{nice04:a}, and
the companion of PSR J1756-2251, with a mass of $1.18^{+0.03}_{-0.02}$
\Msun \cite{faulkner05:a}. It is not clear if the mass range of 1.25
to 1.44 \Msun also applies to {\em isolated} neutron stars.
Evolutionary models of massive stars \cite{TWW96} predict a bimodal
distribution for the neutron star masses, with the two peaks being
between $1.2-1.3$ \Msun for progenitor main sequences masses below 19
\Msun (which will burn carbon in a convective core) and between
$1.7-1.8$ \Msun for progenitor main sequence masses above 19 \Msun
(where carbon burning will occur in a radiative core).  The actual
value of 19 \Msun for the main sequence bifurcation mass may not be
very accurate, but the double peaked neutron star mass distribution is
certainly real.  Given the initial mass function for heavy stars, one
could thus expects that massive (e.g., $> 1.5$ \Msun) isolated neutron
stars be very rare.

A neutron star whose temperature is in disagreement with the Minimal
Cooling predictions must be of course cold, but moreover young and
very likely massive, hence probably uncommon. Therefore, such a star can
be expected to be found far away from the Sun and its detection will
be very difficult since it's weak thermal spectrum will be strongly
absorbed.  Notice that all young ($<$ a few times $10^4$ yrs) neutron
stars plotted in Figure~\ref{fig:minimal} are at a distance superior to
1 kpc except for the Vela pulsar.  In particular, the two most
conspicuous young and cold objects, PSR J0205+6449 (in SNR 3C58) and
RX J0007.0+7302 (in SNR CTA 1) are at distances of $2.6 \pm 0.3$ kpc
and $1.4 \pm 0.3$ kpc, respectively.  Moreover, as emphasized recently
in \cite{Petal03}, about 80\% of the nearby young neutron stars come
from the Gould Belt: this local structure of young stellar
associations forms a ring of about 1 kpc surrounding the Sun and has an
estimated age of 30-50 Myrs.  If the Gould Belt's progenitors of the
nearby young cooling neutron stars were formed when the belt was
formed, which is likely for many of them, and core collapsed about one
Myrs ago, they were massive stars not much above 10 \Msun since heavier
stars died well before.  From the results of \cite{TWW96} one can
safely conclude that most nearby young cooling neutrons stars have masses
not exceeding 1.4 \Msunend: it is not surprising that none of them shows
clear evidence of enhanced neutrino cooling.

The smoking guns in favor of enhanced neutrino cooling are certainly, to
date, the two isolated pulsars PSR J0205+6449 and RX J0007.0+7302 
(see \S~\ref{sec:minimal}) and
the accreting pulsar SAX J1808.4-3658 (see \S~\ref{sec:SXRT}).  Since
evolutionary scenarios for recycling pulsar to millisecond periods
require accretion of about 0.2 to 0.5 \Msun \cite{CST94}, SAX J1808.4-3658
is a natural candidate for a neutron star with an inner core beyond
minimal.
Possibly more such objects will be found in the SNRs studied recently
in \cite{Ketal2004} 
(whose present luminosity upper limit are marked as 
``a'', ``b'', ``c'', and ``d'' in Figure~\ref{fig:minimal}) 
in case a neutron star is detected and will provide us with definite evidence 
for the occurrence of enhanced neutrino emission.
Finally, one must entertain the idea that core collapse supernovae
fail to produce heavy neutron stars, sending them into a black hole 
despite of them being allowed by the equation of state, and that heavy neutron
stars can only be found in binary systems, such as SAX J1808.4-3658, after
substantial mass accretion occurred.

\bibliographystyle{unsrt}
\bibliography{rlist}

\end{document}